\documentclass[amsmath,amssymb,twocolumn,floatfix,pre]{revtex4}

\usepackage{amsmath}
\usepackage{graphicx}

\usepackage[all]{xy}
\usepackage[section]{placeins}
 \,
 \,
 \,
 \,

\renewcommand{\vec}[1]{\mbox{\boldmath $#1$}}

\newcommand{\g}{\vec{\Gamma}}

\begin{document}

\title{Time-reversal symmetry and covariant Lyapunov vectors
for simple particle models in and out of thermal equilibrium}

\author{Hadrien Bosetti}
\email{Hadrien.Bosetti@univie.ac.at}
\affiliation{Computational Physics Group, Faculty of Physics, University of Vienna,
 Boltzmanngasse 5, A-1090 Wien, Austria}
\author{Harald A. Posch}
\email{Harald.Posch@univie.ac.at}
\affiliation{Computational Physics Group, Faculty of Physics, University of Vienna,
 Boltzmanngasse 5, A-1090 Wien, Austria}
\author{Christoph Dellago}
\email{Christoph.Dellago@univie.ac.at}
\affiliation{Computational Physics Group, Faculty of Physics, University of Vienna,
 Boltzmanngasse 5, A-1090 Wien, Austria}
\author{William G. Hoover}
\email{hooverwilliam@yahoo.com}
\affiliation{Ruby Valley Research Institute, Highway Contract 60, Box 598, 
Ruby Valley 89833, NV USA}

\date{\today}

\begin{abstract}
     Recently, a new algorithm for the computation of covariant Lyapunov vectors and of
corresponding local Lyapunov exponents has become available. Here we study the 
properties of these still unfamiliar quantities for a number of simple models, including 
an harmonic oscillator coupled to a thermal gradient with a two-stage thermostat, which
leaves the system ergodic and fully time reversible. We explicitly demonstrate how
time-reversal invariance affects the perturbation vectors  in tangent space and the associated local
Lyapunov exponents. We also find that the local covariant exponents
vary discontinuously along directions transverse to the phase flow.       
\end{abstract}
\maketitle
\section{Introduction}

     Recently, many concepts and methods of dynamical systems  theory have turned out to be
very useful  for the characterization and understanding of  physical systems in and out of
thermodynamic equilibrium. For example, for a class of stationary nonequilibrium systems, the  spectrum 
of Lyapunov exponents is a convenient tool for studying the collapse of the phase-space probability distribution onto fractal measures with an information dimension smaller than the dimension of phase space.
In this case, stationarity is achieved with time-reversible thermostats \cite{HHP87,Hoover_Buch}.
Stationary nonequilibrium systems with stochastic thermostats may be formulated along similar
lines \cite{PH06}.

   The aim of this paper is to apply the hitherto rather unfamiliar concept of covariant  Lyapunov vectors
and their associated local Lyapunov exponents   
to some simple and pedagogical systems in equilibrium and in non-equilibrium stationary states  to
sharpen the intuition for more demanding applications. The systems studied include a
harmonic oscillator subjected to a two-stage chain of Nos\'e-Hoover-type
thermostats with a  temperature which varies with the position of the particle.

   The paper is organized as follows: In the next section we provide the basic theoretical
concepts and definitions required for our numerical work. In particular, the
covariant vectors and their classical counterparts, the Gram-Schmidt vectors, are introduced,
and their dynamical evolution is discussed. Section \ref{Section_differential} is devoted to 
an alternative differential-equation based  method for the evolution of
orthonormal perturbation vectors, which may be interpreted as continuous
re-orthonormalization.  In Section \ref{numerics}  we specify the protocol
for our numerical work, both forward and backward in time. Our main example,
a doubly-thermostated oscillator in a space-dependent thermal field,  is treated in 
various subsections of Section \ref{doubly}.  Section \ref{sym_systems} is devoted to
symplectic systems, with regular trajectories on a torus or with chaotic behavior, for which the 
differences of the symmetry properties for the local Gram-Schmidt and covariant Lyapunov exponents 
are most pronounced. We conclude in Section \ref{conclude} with some remarks, which also
concern the stationary fluctuation theorem for thermostated systems.   
 

\section{Covariant Lyapunov vectors and local Lyapunov exponents}
\label{instant}
If ${\bf \Gamma}(t)$ denotes the state of a dynamical system of dimension $D$,
its evolution equations are given by
\begin{equation}
    \dot{\bf \Gamma} = {\bf F}({\bf \Gamma}),
   \label{motion} 
\end{equation}     
where ${\vec F}$ is a (generally nonlinear) vector-valued function of dimension $D$.
An arbitrary perturbation vector $\delta {\bf \Gamma}(t)$ in tangent space evolves according 
to the linearized equations
\begin{equation}
    \dot{\delta {\bf \Gamma}} =  {\cal J}({\bf \Gamma})  \delta {\bf \Gamma},
\label{linearized}
\end{equation}    
where the dynamical (Jacobian) matrix ${\cal J}$ is given by
$$
{\cal J}({\bf \Gamma}) = \frac{ \partial {\vec F}}{ \partial {\bf  \Gamma}}.
$$ 
The stability of a trajectory in a $D$-dimensional phase space is determined by a 
set of $D$ (global) Lyapunov exponents, which are the {\em time-averaged} logarithmic rates of the growth or
decay of the norm of some perturbation vectors, which must be  oriented `properly' in tangent space at 
the initial time.  Formally, let ${\bf \Gamma}(0)$ denote the state of the system at time $0$,
the state at time $t$ is given by  ${\bf \Gamma}(t) =  \phi^t ({\bf \Gamma}(0))$, 
where  the map $\phi^t:  {\bf \Gamma} \to {\bf \Gamma}$ defines the flow in the phase space ${\bf \Gamma}$.
Similarly, if  $\delta{\bf \Gamma}(0)$ is a vector  in the tangent space  at the phase point
${\bf \Gamma}(0)$, at time $t$, it becomes $\delta{\bf \Gamma}(t) =  D\phi^t \vert_{{\bf \Gamma}(0)}\;
\delta{\bf \Gamma}(0)$, where $D\phi^t$ defines the tangent flow.  It is represented by a
real but generally nonsymmetric $D \times D$ matrix. The multiplicative ergodic theorem of  Oseledec
\cite{Oseledec:1968,Ruelle:1979,Eckmann:1985}
 asserts that there exist `properly oriented'  and normalized vectors 
${\vec v}^{\ell}\left({\bf \Gamma}(0)\right)$ in tangent space at $t = 0$, which evolve according to 
\begin{equation}
 D \phi^t \vert_{{\bf \Gamma}(0)} \; \vec v^{\ell}\left({\bf \Gamma}(0)\right) = \vec v^{\ell}\left( {\bf  \Gamma}(t)\right) , 
\label{defcov}
\end{equation}
and which generate the Lyapunov exponents on the way, 
\begin{equation}
 \pm \lambda_{\ell}   = \lim_{t \rightarrow \pm \infty} \dfrac{1}{\vert t \vert} \, \ln \, \big\| \, D\phi^t\vert_{\textrm{\smallskip{\g}(0)}} \, \; \, {\vec v}^{\ell}\left({\bf \Gamma}(0)\right) \, \big\|
\label{covlambda}
\end{equation}
for all  $\ell \in \{ 1,\ldots, D \}$, both forward and backward in time (for time-reversible systems). (Strictly speaking, this formulation is only correct for nondegenerate exponents $\lambda_{\ell}$. If two such exponents become
identical, the respective vectors must be replaced by a covariant subspace spanned by the vectors. Since
in our applications below, this happens only for the symplectic systems in thermodynamic equilibrium
discussed in Sec. \ref{sym_systems}, there is no danger of misinterpretation, and we avoid the additional 
notational complexity. The case of degenerate exponents is treated in detail in Ref. \cite{BP10}). 
Because of the property described by Eq. (\ref{defcov}), the vectors ${\vec  v}^{\ell}$ are called {\em covariant}.
Loosely speaking, covariant vectors are co-moving (co-rotating in particular)
with the tangent flow. As will be shown below, this property of co-rotation is responsible for the fact that
the evolution of their length in the forward and backward directions of time (for time-reversible systems)
is intimately connected,  a symmetry not enjoyed by other perturbation vectors.  
For numerical  reasons, it is still necessary to {\em normalize} the vectors periodically at times 
$t_n \equiv n \tau$, such that Eq. (\ref{covlambda}) becomes
\begin{equation}
 \lambda_{\ell} = \lim_{N \rightarrow  \infty} \frac{1}{N\tau} \sum_{n=0}^{N-1}
     \ln \big\|  D \phi^{\tau} \vert_{{\bf \Gamma}_n} \; {\vec v}^{\ell}({\bf \Gamma}_n )\big\|,
     \label{intcov}
\end{equation}     
    where we use the abbreviated notation ${\bf \Gamma}(t_n) \equiv {\bf \Gamma}_n$.
$ \| {\vec v}^{\ell}({\bf \Gamma}_n )\| = 1$ at the beginning of each interval of length $\tau$.
Generally its norm differs from unity at the end of the interval.

Up to very recently, no practical algorithm for the computation of the covariant vectors was available.
The classical algorithm  for the computation of Lyapunov exponents \cite{Benettin,Shimada} is based 
on the fact that almost all volume elements of dimension $d \le D$ in tangent space (with the exception of elements of measure zero)  asymptotically evolve  with an exponential rate, which is equal to the sum of the 
first $d$ Lyapunov exponents.  Such a  $d$-dimensional subspace may be spanned by 
$d$ orthonormal vectors, which are constructed by the Gram-Schmidt re-orthonormalization
procedure and, therefore,  are referred to as Gram-Schmidt (GS) vectors ${\vec g}^{\ell}({\bf \Gamma}(t))$. 
The evolution during the time interval $\tau = t_n - t_{n-1}$,
\begin{equation}
 D \phi^{\tau} \vert_{{\bf \Gamma}_{n-1}} \; \vec g^{\ell}\left({\bf \Gamma}_{n-1}\right) \equiv 
 \bar{\vec g}^{\ell }\left({\g_n}\right), 
\label{defgs}
\end{equation}
generates a set of non-orthonormal vectors , $\{ \bar{\vec g}^{\ell }\left({\g_n}\right), \; \ell = 1,\cdots,D\}$,
which after Gram-Schmidt re-ortho\-normalization \cite{Wolf,recipes},
 \begin{eqnarray}
 {\vec g}^{1 }\left({\g_n}\right) &= &\frac{ \bar{\vec g}^{1 }\left({\g_n}\right)}{\big\|
        \bar{\vec g}^{1 }\left({\g_n}\right) \big\|}, \nonumber \\
 {\vec g}^{\ell }\left({\g_n}\right) &= &\frac{ \bar{\vec g}^{\ell }\left({\g_n}\right) -\sum_{k=1}^{\ell-1} 
       \left(  \bar{\vec g}^{\ell }\left({\g_n}\right)     \cdot     {\vec g}^{k }\left({\g_n}\right)   \right) {\vec g}^{k }\left({\g_n}\right)} 
{ \big\|
\bar{\vec g}^{\ell }\left({\g_n}\right) -\sum_{k=1}^{\ell-1} 
       \left(  \bar{\vec g}^{\ell }\left({\g_n}\right)     \cdot     {\vec g}^{k }\left({\g_n}\right)   \right){\vec g}^{k }\left({\g_n}
 \right) \big\|} \nonumber,
\end{eqnarray}
  (where $\ell$ consecutively assumes the values  $1, \cdots,D$) become the orthonormal starting vectors for the next interval.
 The vectors ${\vec g}^{\ell}$ are not covariant,
which means that, in general,   the vectors are not mapped 
by  the linearized dynamics into the GS vectors at the forward images of the initial phase-space  
point \cite{Ginelli}. As a consequence, they are also not invariant with respect to the time-reversed dynamics. 
The Lyapunov exponents are computed from the normalization factors,
\begin{eqnarray}
\lambda_1       & = & \lim_{N \rightarrow  \infty} \frac{1}{N\tau} \sum_{n=0}^{N-1}\ln    \big\|
\bar{\vec g}^{1 }\left({\g_n}\right) \big\|,   \nonumber \\
\lambda_{\ell} & = & \lim_{N \rightarrow  \infty} \frac{1}{N\tau} \sum_{n=0}^{N-1}
   \ln    \big\|
\bar{\vec g}^{\ell }\left({\g_n}\right)  \nonumber \\
   & &-\sum_{k=1}^{\ell-1} 
       \left(  \bar{\vec g}^{\ell }\left({\g_n}\right)     \cdot     {\vec g}^{k }\left({\g_n}\right)   \right){\vec g}^{k }\left({\g_n}
 \right) \big\|   \label{intgs}
\end{eqnarray} 
for $\ell = 2,\cdots,D$.    

Recently, a reasonably fast algorithm for the computation of covariant Lyapunov vectors
was presented by Ginelli {\em et al.}  \cite{Ginelli}, which first requires the  construction  of the
Gram-Schmidt vectors by a forward integration in time. In a second step, this stored information 
is used to obtain the covariant vectors by a backward iteration in time. For details of this 
algorithm we refer to their paper \cite{Ginelli} and to our previous work \cite{BP10}. 

A  {\em local}  Lyapunov exponent characterizes the expansion, or shrinkage,
of a particular tangent vector  during a (short) time interval $\tau$.  From
Eqs. (\ref{intcov}) and (\ref{intgs})  local exponents for the covariant and
Gram-Schmidt vectors are obtained for a time $t_n \equiv n \tau$ at the phase point ${\bf \Gamma}_n$:
\begin{equation}
\Lambda_{\ell}^{\mbox{cov}}(t_n) =   \frac{1}{\tau}
     \ln \big\|  D \phi^{\tau} \vert_{{\bf \Gamma}_{n-1}} \; {\vec v}^{\ell}({\bf \Gamma}_{n-1} )\big\|,
     \label{loccov} 
\end{equation}
for $\ell = 1,\cdots,D$, and
\begin{eqnarray}     
\Lambda_1^{\mbox{GS}}(t_n)      & = &  \frac{1}{\tau} \ln    \big\|
\bar{\vec g}^{1 }\left({\g_n}\right) \big\|,   \nonumber \\
\Lambda_{\ell}^{\mbox{GS}}(t_n)  & = &  \frac{1}{\tau}
   \ln    \big\| \bar{\vec g}^{\ell }\left({\g_n}\right)  \nonumber \\
   & &-\sum_{k=1}^{\ell-1} 
       \left(  \bar{\vec g}^{\ell }\left({\g_n}\right)     \cdot     {\vec g}^{k }\left({\g_n}\right)   \right){\vec g}^{k }\left({\g_n}
 \right) \big\|,   
     \label{locGS}
\end{eqnarray}
for $\ell = 2,\cdots,D$.

Since the spaces 
\begin{equation}
{\vec v}^1 \oplus \cdots \oplus {\vec v}^{\ell}  = {\vec g}^1 \oplus \cdots \oplus {\vec g}^{\ell}  
\label{subspaces}
\end{equation}
are covariant subspaces of the tangent space for all $ \ell$, 
we have ${\vec v}^{\ell}(t) \in {\vec g}^1(t) \oplus \cdots \oplus {\vec g}^{\ell}(t)$.
If $\beta_{\ell \ell}(t) $  denotes the 
angle between the respective covariant and  Gram-Schmidt vectors ${\vec v}^{\ell}(t)$ and
${\vec g}^{\ell}(t)$  at the specified time,
the component of the normalized vector ${\vec v}^{\ell}(t_{n-1})$ in the direction of 
${\vec g}^{\ell}(t_{n-1})$ is given by
$\cos \beta_{\ell \ell }(t_{n-1})$. During $\tau$, this vector component grows by a factor 
$\exp\{ \Lambda_{\ell}^{\mbox{GS}} \tau\}$, 
whereas the norm of the vector itself grows by   $\exp\{ \Lambda_{\ell}^{\mbox{cov}} \tau \} $.
At the end of the interval, equating the vector components of  ${\vec v}^{\ell}(t_n)$  in the {\em new} direction of 
the re-orthonormalized vector ${\vec g}^{\ell}(t_n)$, one obtains
\begin{equation}
     \Lambda_{\ell}^{\mbox{GS}}(t_n) = \Lambda_{\ell}^{\mbox{cov}}(t_n) + \frac{1}{\tau} \ln \frac{\cos \beta_{\ell \ell}(t_n)} {\cos \beta_{\ell \ell}(t_{n-1})}, 
\label{inst_lya}   
\end{equation}  
$ \ell  = 1,\cdots, D$. This relates the local exponents  for the GS and covariant vectors. 
 
      If we consider the limit $\tau \to 0$ implying continuous re-orthonormalization of the  ${\vec g}^{\ell}$
and normalization  of the ${\vec v}^{\ell}$,      Eq. (\ref{inst_lya}) becomes
$$
     \Lambda_{\ell}^{\mbox{GS}}(t) = \Lambda_{\ell}^{\mbox{cov}}(t) -
             \tan \beta_{\ell \ell}(t) \frac{ d \beta_{\ell \ell}}{ d t}  .
$$
This is most easily achieved with a matrix of Lagrange multipliers constraining the vectors to unit length
and enforcing orthogonality of the    ${\vec g}^{\ell}$ \cite{Goldhirsch,PH88,PH04}. We shall return
to this point in the following section.
  
For time-continuous systems, these relations are general and are not restricted to any particular model.
Below they will be applied to a variety of models mentioned in the introduction.

The global exponents are the time averages of the local exponents over a long trajectory tracing out
the whole ergodic phase space component, and are the same for the covariant and Gram-Schmidt cases,
$$
\lambda_{\ell}  = \lim_{N \to \infty} \frac{1}{N}  \sum_{n=0}^{N-1}   \Lambda_{\ell}^{\mbox{cov}}(t_n) = 
              \lim_{N \to \infty} \frac{1}{N}  \sum_{n=0}^{N-1}   \Lambda_{\ell}^{\mbox{GS}}(t_n).
$$
Whereas the global Lyapunov exponents do not
depend on the particular  metric and the choice  of the coordinate system,  the
local exponents do. This will become apparent in Section \ref{sym_systems} for
a scaled harmonic oscillator. For particular applications of the local exponents this must be kept 
in mind.

   All systems we consider here are invariant with respect to time  reversal. This property leaves the 
equations of motion in phase and tangent space unchanged if the signs of all momentum-like variables 
{\em and} of time are reversed, but leaving all position variables unchanged. This implies 
that there exists a smooth isometry $I$ of phase space, such that $ I \phi^t = \phi^{-t} I $.
In practice, an  integration of the equations of motion backward in time is carried out with reversed 
momentum-like variables and a positive time step. After reaching the endpoint, the signs of all
momentum-like variables need to be reversed again and the time variable properly adjusted.
 Alternatively, and even more easily,
the integration of the motion equations may proceed without changing the sign of the
momentum-like variables but with a negative time step. There is also no sign change after reaching
the end point in this case. A comparison of both methods yields  identical results. 
Where necessary, we indicate the forward and backward directions of
time by upper indexes $(+)$ and $(-)$, respectively. If this index is omitted, the forward direction is implied. 

We have mentioned already that the classical algorithm invoking Gram-Schmidt re-orthonormalization
carefully keeps track of the time evolution of $d$-dimensional volume elements, $\delta {\cal V}_d$,
for any $d \le D$, which proceeds according to  \cite{Hoover_Buch,PDHK97}
$$
        \frac{ d \ln \delta {\cal V}_d (t)}{dt}  =  
         \sum_{\ell = 1}^d \Lambda_{\ell}^{\mbox{GS}}(t).
 $$        
If the total phase volume is conserved as for symplectic systems, the following sum rule 
for the Gram-Schmidt local exponents holds at all times:
\begin{equation}
            \sum_{\ell = 1}^D \Lambda_{\ell}^{\mbox{GS}}(t) = 0.
\label{sum_rule}
\end{equation}
In this symplectic case we can even say more. For each positive local GS exponent there is a 
local negative GS exponent such that their pair sum vanishes \cite{Mayer}, 
\begin{eqnarray}
^{(+)}\Lambda_{\ell}^{\mbox{GS}}(t) &=&  -^{(+)}\Lambda_{D + 1 -\ell }^{\mbox{GS}}(t), \label{gsfwd} \\ 
^{(-)}\Lambda_{\ell}^{\mbox{GS}}(t) &=&  -^{(-)}\Lambda_{D + 1 -\ell }^{\mbox{GS}}(t). \label{gsbwd} \\ 
\nonumber
\end{eqnarray}
As indicated, such a symplectic local pairing symmetry exists both forward and backward in time.
But, generally, the GS local exponents do not show the symmetry with respect to time-reversal invariance. Thus,
\begin{equation}
^{(-)}\Lambda_{\ell}^{\mbox{GS}}(t) \ne -^{(+)}\Lambda_{D + 1 -\ell }^{\mbox{GS}}(t). 
\label{not_equal}
\end{equation}
No such symmetries exist for non-symplectic systems. Examples are provided below. 
 
The situation is very different for the covariant local Lyapunov exponents. In their case, the vectors 
are still re-normalized,  but the angles between them remain unchanged, which effectively destroys 
all information concerning the $d$-dimensional volume elements. Thus, no symmetries 
analogous to Eqs. (\ref{gsfwd}) and ({\ref{gsbwd}) exist.  Instead, the re-normalized covariant vectors faithfully preserve the time-reversal symmetry of the equations of motion, which is  reflected by
\begin{equation}   
 ^{(-)}\Lambda_{\ell}^{\mbox{cov}}(t)  = -^{(+)}\Lambda_{D+1 -\ell}^{\mbox{cov}}(t) \; \mbox{for} \;  \ell = 1,\cdots,D,
 \label{local_symmetry}
\end{equation}
regardless, whether the system is symplectic or not. This means that an expanding co-moving direction is
converted into a collapsing co-moving direction by an application of the time-reversal operation.
Of course, the forward and backward local exponents in Eq. \ref{local_symmetry})  refer to the
same phase space point ${\bf \Gamma}(t)$. 

These symmetry properties may be considered the main
conceptual differences between the Gram-Schmidt and covariant viewpoints.  

Before leaving this section, a short remark concerning the commonly-used term
``time-dependent exponent'' seems in order. Primarily, this quantity is a function of the
phase point and should only be called a ``local'' exponent.  Its value may be different
whether the phase point is reached from the past, forward in time $(+)$, or from the future,
backward in time $(-)$.

\section{Differential approach to local Lyapunov exponents}
\label{Section_differential}

   Equation (\ref{locGS}) precisely reflects the numerical procedure for the computation of 
local GS exponents for finite time intervals $\tau$. But it is also possible to
obtain a differential version for $\tau \to 0$.   Goldhirsch {\em et al.} derived a full set of differential equations 
for the Gram-Schmidt vectors ${\vec g}^{\ell}$ \cite{Goldhirsch},
\begin{eqnarray}
\dot{\vec g}^1 & = & {\cal J} {\vec g}^1  - R_{11}\; {\vec g}^1, \label{l1} \\ 
\dot{\vec g}^{\ell} & = & {\cal J} {\vec g}^{\ell}  - R_{\ell \ell}\;  {\vec g}^{\ell}  - \sum_{m=1}^{\ell-1} 
(R_{\ell m} + R_{m \ell}) \; {\vec g}^m, \label{l2}\\
\nonumber
\end{eqnarray} 
where in the last equation $\ell = 2, \cdots, D$.  We have demonstrated  \cite{PH88,PH04} that
the matrix elements  
\begin{equation}
R_{\ell m}\left({\bf \Gamma}(t)\right) = ({\vec g}^{\ell})^T  {\cal J}  {\vec g}^m
\end{equation} 
 may be understood as Lagrange multipliers enforcing the
orthonormalization constraints
$ {\vec g}^{\ell} \cdot {\vec g}^m = \delta_{\ell m}$ (equal to unity for $\ell = m$, and  zero
otherwise). Here $T$ means transposition. Most importantly,    
the diagonal elements  are the local Gram-Schmidt exponents:
\begin{equation}
    \Lambda_{\ell}^{\mbox{GS}}({\bf \Gamma}(t)) \equiv R_{\ell \ell}({\bf \Gamma}(t)) = 
    ({\vec g}^{\ell})^T   {\cal J}  {\vec g}^{\ell}.
 \label{continb}
 \end{equation}   
 This expression nicely underlines the local nature of the exponents. 

We have verified for the doubly-thermostated heat conduction model
discussed in Sec. \ref{doubly} below that the direct integration of the Eqs. (\ref{l1},\ref{l2}) provide
local GS exponents according to Eq. (\ref{continb}), which agree extremely well with the 
results obtained from a direct application of the GS algorithm,  
Eq. (\ref{locGS}), for a reasonably-small time interval $\tau$. 
This agreement also persists for the time-reversed dynamics.

\section{Numerical considerations} 
\label{numerics}


In this section we remark on a few aspects of our  implementation of the   algorithm 
for the  computation of the covariant Lyapunov vectors, which we apply in the following sections. 
Reduced units are used for the various models treated  below. For convenience, we specify
already here the adopted values (in reduced units) for some time parameters: 
$t_{\omega} = 6 \times 10^4$, $t_{\alpha} = 5\times 10^4$, and
$t_0 = 100$.  Their meaning is explained below. For the integration of the equations of motion, a 
4th-order Runge-Kutta algorithm with a time step $dt = 0.001$ is used. This time step is chosen
such that the trajectory is correct to double-precision accuracy. For the interval between
successive Gram-Schmidt re-orthonormalization steps -- respective covariant vector normalizations -- 
we choose $\tau = 10 dt = 0.01$. This number is a (very conservative)  compromise between the achieved 
reduction in storage requirements as outlined below, and the precision of integration forward and
backward over the same interval. The time $t_0$ is chosen such that in the interval $-t_0 \le t \le t_0$
accurate Gram-Schmidt and covariant Lyapunov vectors are available.

The simulations are carried out with the following protocol:

\noindent
{\bf Phase 1 (forward integration from $-t_{\omega}$ to $ +t_{\omega}$)}: Starting with  arbitrary initial conditions 
at a time $-t_{\omega} $ and using a positive integration time step  $dt  > 0$,  
the evolution of the reference trajectory ${\vec \Gamma(t)}$ and of the full set of Gram-Schmidt vectors is
computed in the forward direction of time up to a time $+t_{\omega} $.
The reference trajectory and the Gram-Schmidt vectors are stored for every 
10 time steps, $10 dt = \tau$, along the way. The Gram-Schmidt vectors are used in phase 2  for the
construction of the covariant vectors, and the reference trajectory is required in phase 3 for the
computation of  the time-reversed Gram-Schmidt vectors.   The Lyapunov spectrum 
$  \{ ^{(+)}\lambda^{\mbox{GS}} \}$ is accumulated for times
$-t_{\alpha} \le t \le  t_{\omega}$, for which  the orientations of the Gram-Schmidt vectors are fully relaxed.  

\noindent 
 {\bf Phase 2 (backward iteration from $t_{\omega} $ to $- t_0$)}: 
 Starting at $t_{\omega}$, the covariant 
 vectors are computed by iterating back to a time $-t_0$. The details of this algorithm
 are given in Ref. \cite{BP10}. Since the {\em forward} GS-vectors, stored during phase 1,
 are now used in reversed order,  the consecutive order of the covariant vectors 
 $ \cdots, {\vec v}^{\ell}(t_n),   {\vec v}^{\ell}(t_{n-1}), \cdots$ has to be reversed for the 
 computation of the corresponding local exponent, 
 $$^{(+)}\Lambda_{\ell}^{\mbox{cov}}( t_n) 
 = \frac{1}{\tau} \ln \frac{ \|{\vec v}^{\ell}(t_n)\| }{ \|{\vec v}^{\ell}(t_{n-1})\|},$$
 or, alternatively, the sign of the local exponents must be reversed. The time averaging for the 
 global Lyapunov spectrum $ \{^{(+)}\lambda_{\ell}^{\mbox{cov}} \} $
 is carried out for times  $t_{\alpha} \ge t \ge - t_0$.

The following two  phases are only required for the computation of the local
{\em time-reversed} Gram-Schmidt and covariant exponents.  

\noindent
{\bf Phase 3 (backward integration from  $t_{\omega}$ to $ -t_{\omega}$)}:
With arbitrary initial conditions  at time $t_{\omega}$,
the Gram-Schmidt tangent space dynamics is  followed  backward in time up to $-t_{\omega}$.
To counteract the Lyapunov instability, it is essential for this computation 
to use the {\em same} reference trajectory stored in phase 1, where the sign of the     
momentum-like variables (  $p, z,$ and $x$ for the doubly-thermostated oscillator) 
is left unchanged, but with the  time step reversed to $-0.001$. For an accurate
 computation of the backward GS vectors, the reference trajectory at {\em every} integration step
 is required. Since in phase 1 this information was stored for only every 10th step  (to save
 computer storage), the forward reference trajectory  is piece-wise re-computed with
 stored  phase-space points as initial conditions. The backward Gram-Schmidt vectors are
 again stored for every 10th time step replacing the forward vectors of phase 1. 
 If time is reversed, the stable directions  become unstable and {\em vice versa}.
 The  Lyapunov spectrum  $  \{^{(-)}\lambda_{\ell}^{\mbox{GS}} \} $  is accumulated in the interval 
 $t_{\alpha} \ge t \ge -t_{\omega}$.

\noindent 
{\bf Phase 4 (forward iteration from $-t_{\omega }$ to $+t_0$)}: 
Analogous to phase 2, in this final stage the covariant Lyapunov vectors for the 
{\em reversed time direction} are computed with the help of the time-reversed
Gram-Schmidt vectors from phase 3. The respective Lyapunov spectrum 
$ \{^{(-)}\lambda_{\ell}^{\mbox{cov}}\} $ is  accumulated for times  $-t_{\alpha} \le t \le t_0$.

It may be noticed that in the interval $-t_0 \le t \le +t_0$ all local properties are available with
the Gram-Schmidt and covariant vectors fully relaxed both forward and backward in time. 
Therefore, the detailed analysis of local (time-dependent) Lyapunov exponents in the following 
sections is carried out in this regime.


\section{Doubly-thermostated oscillator}
\label{doubly}
\subsection{Description of the model}
\label{model_description}
    Here  we consider a simple model which already  has many ingredients in common with 
much more involved physical systems. It exhibits chaotic equilibrium and stationary nonequilibrium states
and collapses onto a limit cycle for very strong driving.
It consists of a one-dimensional harmonic oscillator, which is  coupled to two consecutive stages
of Nos\'e-Hoover thermostats with a  space-dependent temperature $T(q)$. The equilibrium version
of this model was first considered by Martyna, Klein and Tuckerman \cite{Martyna}. Its nonequilibrium
properties were consecutively studied by us in some detail \cite{HPH01,HHP01}, but without considering covariant vectors. This  paper is also intended to augment this work correspondingly.

The equations of motion, expanded with two thermostat variables $z$ and $x$, are \cite{HPH01,HHP01}
\begin{eqnarray}
\dot{q} &=& p, \nonumber\\
\dot{p} &=& - q - z p, \nonumber\\
\dot{z} &=& p^2 - T(q) - z x,  \nonumber\\
\dot{x} &=& z^2 - T(q), \nonumber\\  
\nonumber
\end{eqnarray}
where the position dependent temperature is given by
$$
     T(q) =  1 +  \varepsilon \tanh (q) .
$$     
The control parameter $\varepsilon$ coincides with the temperature gradient at $q = 0$.
These equations are written in the most simple reduced form with all arbitrary parameters 
of the model set equal to unity. The system is not symplectic.  On average, the oscillator 
picks up energy from the thermostat 
whenever it is in a  region of high temperature ($q > 0$), and  releases it again 
in low-temperature regions ($q < 0$). 

\subsection{Global properties}

For a typical non-equilibrium state, $(\varepsilon = 0.25)$,   the global Lyapunov spectrum 
was computed  by four independent methods, applying the protocol outlined   in Sec. \ref{numerics}: 

\begin{description}
\item[Phase 1]: GS exponents in forward direction of time,  \\
$     \{^{(+)}\lambda_{\ell}^{\mbox{GS}}\}  = \{0.053_1, 0.0000_1, -0.034_4, -0.086_7\} $, \\
 \item[Phase 2]: covariant exponents in forward direction of time \\
 $     \{^{(+)}\lambda_{\ell}^{\mbox{cov}} \} = \{0.053_6, 0.0000_1, -0.035_1, -0.086_2\} $, \\
 \item[Phase 3]:  GS exponents  in backward direction of time\\
 $     \{^{(-)}\lambda_{\ell}^{\mbox{GS}}\}  = \{0.086_7, 0.034_4, 0.0000_3, -0.053_1\} $, \\
\item[Phase 4]: covariant exponents in backward direction of time \\
 $     \{^{(-)}\lambda_{\ell}^{\mbox{cov}}\}  = \{0.087_1,  0.033_7, 0.00000_1, -0.052_5\}. $ \\
 \end{description}
 The last digit of each number is rounded accordingly.
 Considering the smallness of the exponents and  the rather involved numerical procedures, 
 the agreement  between the  independently-determined global spectra is very
 satisfactory. A comparison of the  forward and backward dynamics reveals the
theoretically expected symmetry  for the global Lyapunov exponents \cite{Ruelle:1979,Eckmann:2005},
 \begin{equation}   
 ^{(-)}\lambda_{\ell}  = -^{(+)}\lambda_{D+1 -\ell} \; \mbox{for} \;  \ell = 1,\cdots,D.
 \label{global_exponent}
\end{equation}

If the temperature gradient $\varepsilon$ is varied over a wide range, significant changes of the
spectrum become evident. This is shown in the top panel of Fig. \ref{figure_1}. 
\begin{figure}[ht]
\center
\includegraphics[width=0.5\textwidth]{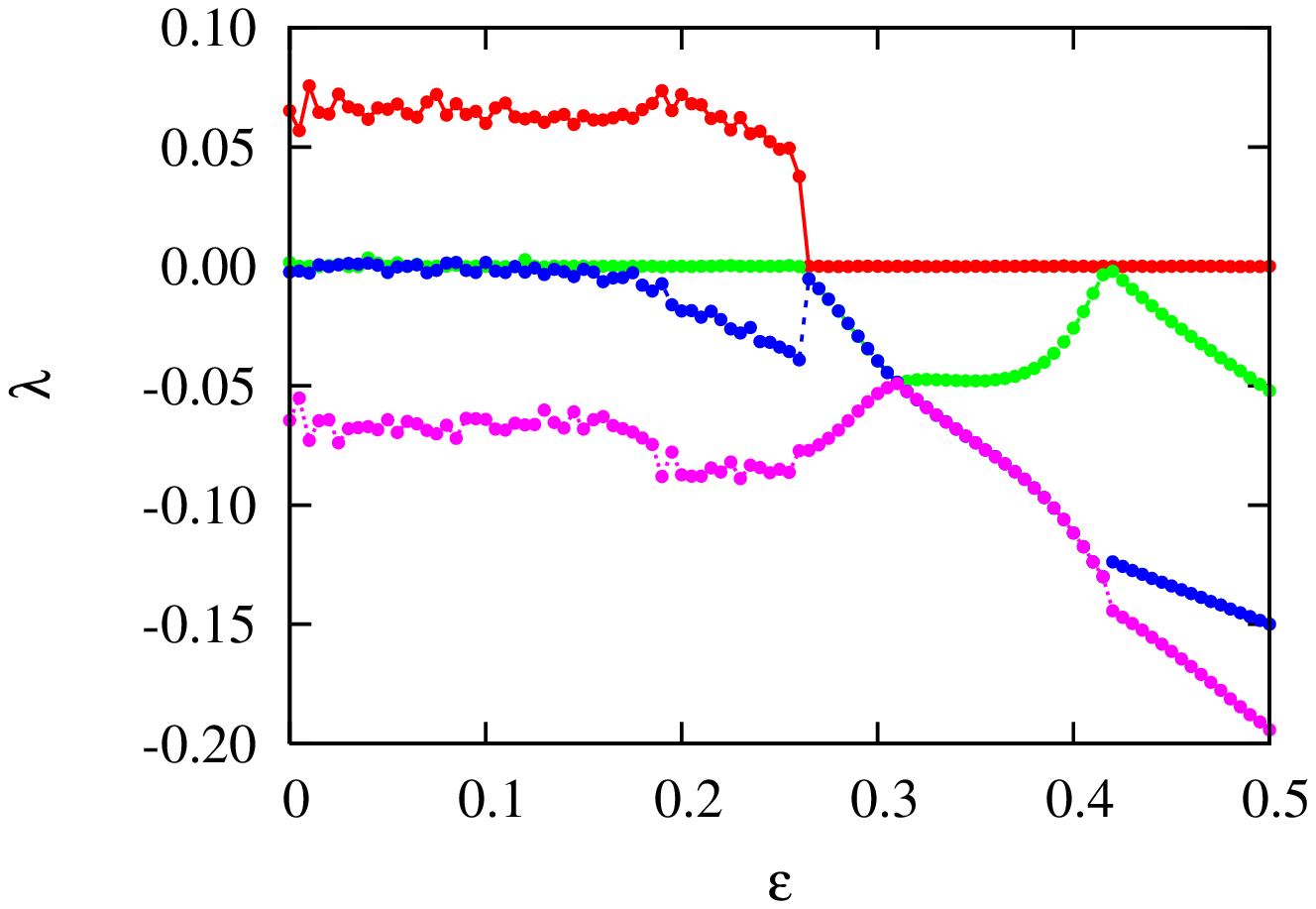}\\
\vspace{-0.5cm}
\includegraphics[width=0.5\textwidth]{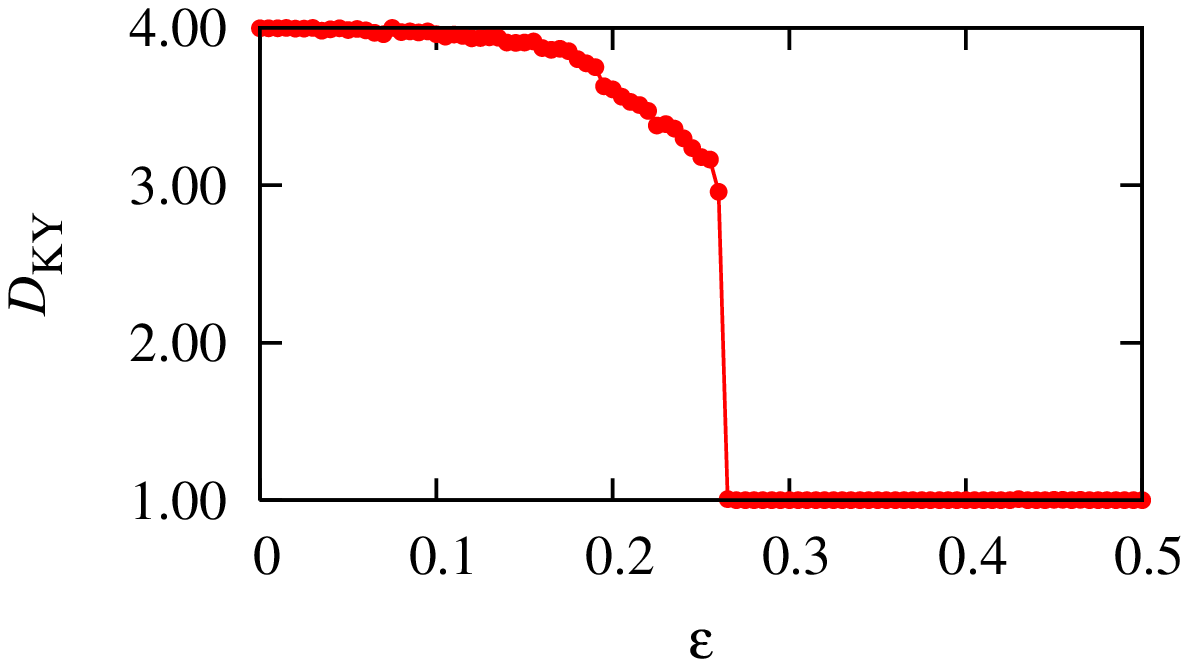}
\vspace{-1cm}
\caption{(Color online) Temperature-gradient dependence of all four Lyapunov  exponents (top panel)
and of the Kaplan-Yorke dimension (bottom panel) 
for the doubly-thermostated oscillator. For $\varepsilon > \varepsilon_c =  0.26314$, the trajectory collapses
onto a limit  cycle.  }
\label{figure_1}
\end{figure}
There exist a number of distinct regimes with different qualitative behavior.

 For $ \varepsilon \lesssim 0.18$,
the spectrum changes but little with $\varepsilon$, and the Kaplan-Yorke dimension is 
only weakly reduced with respect to the full phase space dimension, as is shown in 
the bottom panel of Fig. \ref{figure_1}. The dissipation due to  the weak heat current influences 
the appearance of the chaotic phase-space trajectory very little.  An example 
of a projection of such a trajectory onto the $qpz$-subspace is
provided in the top panel of Fig. \ref{figure_5}.

For $0.18 \lesssim \varepsilon < 0.26$, the trajectory is more and more attracted 
to a weakly unstable periodic orbit  (see the bottom panel of Fig. \ref{figure_5}),
 which for $\varepsilon_c \approx 0.26312$ turns into
a stable limit cycle as shown in the top panel of Fig. \ref{figure_4}.
The nature of this transition may be established by considering the Floquet multipliers
$\mu_{\ell}, \; {\ell} = 1,\cdots,4$ for the fixed points of the Poincar\'e map, defined by
$q= 0$,  for $\varepsilon \ge \varepsilon_c \approx 0.26312$. Whereas ${\mu_1} = 1$
and $\mu_4 < 0$, a single mutiplier $\mu_2$ crosses the unit circle on the real axis at the point A
corresponding to $\varepsilon_c$ in Figure \ref{figure_2}. 
Such a behavior is characteristic of a period doubling bifurcation
\cite{Kuznetsov}, where, possibly, the chaotic attractor disappears in a boundary
crisis bifurcation. This point will be studied separately \cite{HB}.  Increasing $\varepsilon$
further, the Floquet multipliers $\mu_{2,3}$ vary as indicated by the arrows and become
complex conjugate to each other for $\varepsilon \approx 0.26319$ (point B in Fig. \ref{figure_2}).
\begin{figure}[ht]
\center
\includegraphics[width=0.4\textwidth]{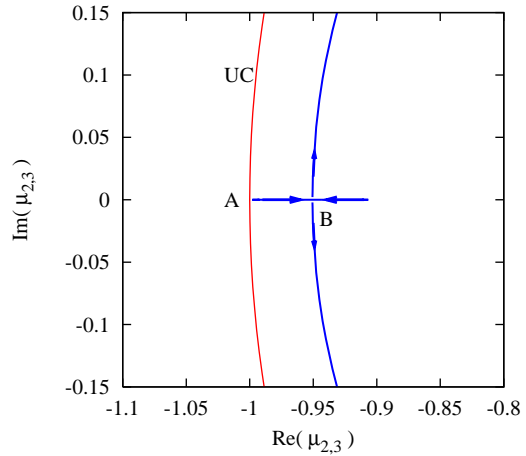}\\
\vspace{-0.5cm}
\caption{(Color online) Floquet multipliers $\mu_2$ and $\mu_3$ in the complex plain for $\varepsilon \ge  0.26312$. UC denotes the unit circle, and  A is the bifurcation point. The multipliers are
complex conjugate to each other for $\varepsilon > 0.26319$ as indicated by  B.
}
\label{figure_2}
\end{figure}

For $ \varepsilon \approx 0.417$, there is another transition changing the two-loop limit cycle
into a single-loop orbit. This is illustrated in the bottom panel of Fig. \ref{figure_4} and 
will be studied separately \cite{HB}.

\begin{figure}[ht]
\center
\includegraphics[width=0.5\textwidth]{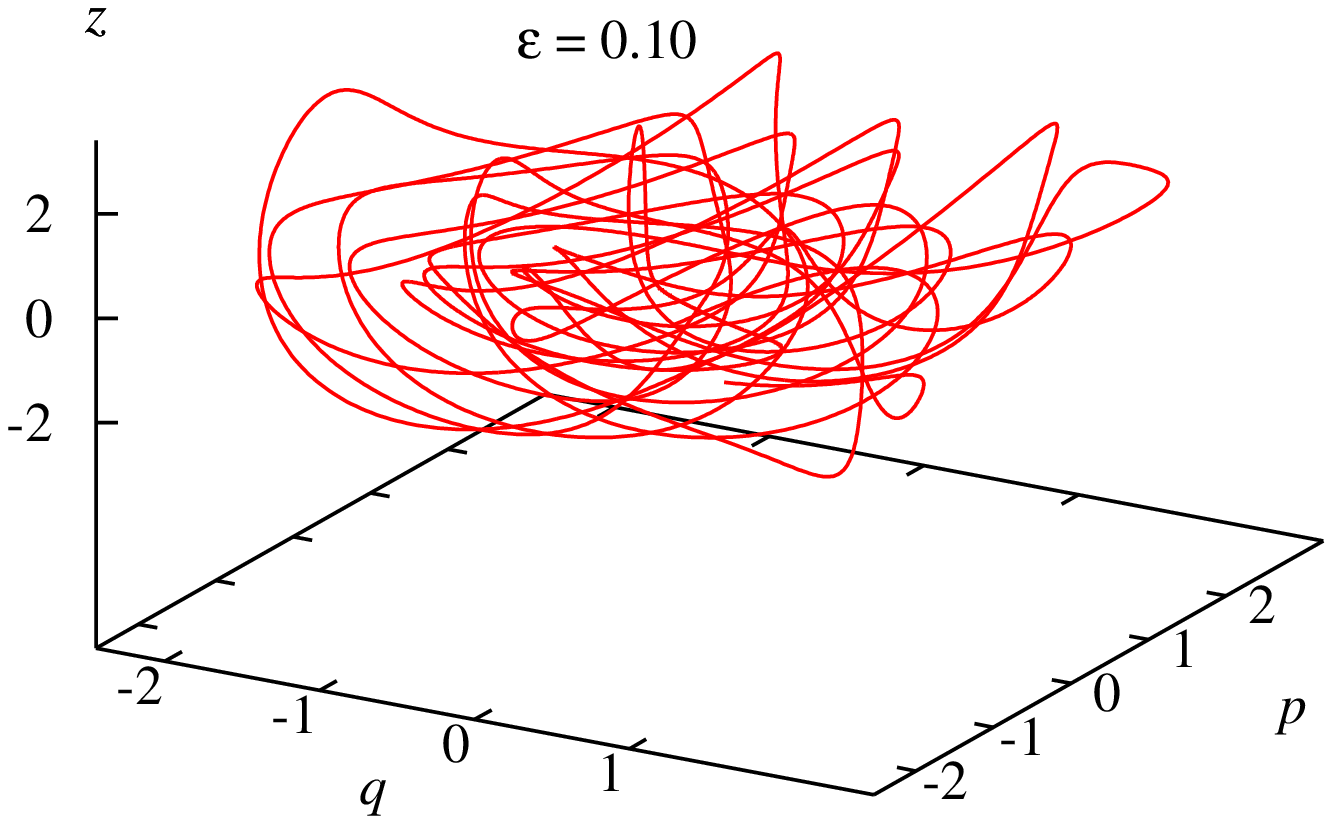}\\
\vspace{-1.0cm}
\includegraphics[width=0.5\textwidth]{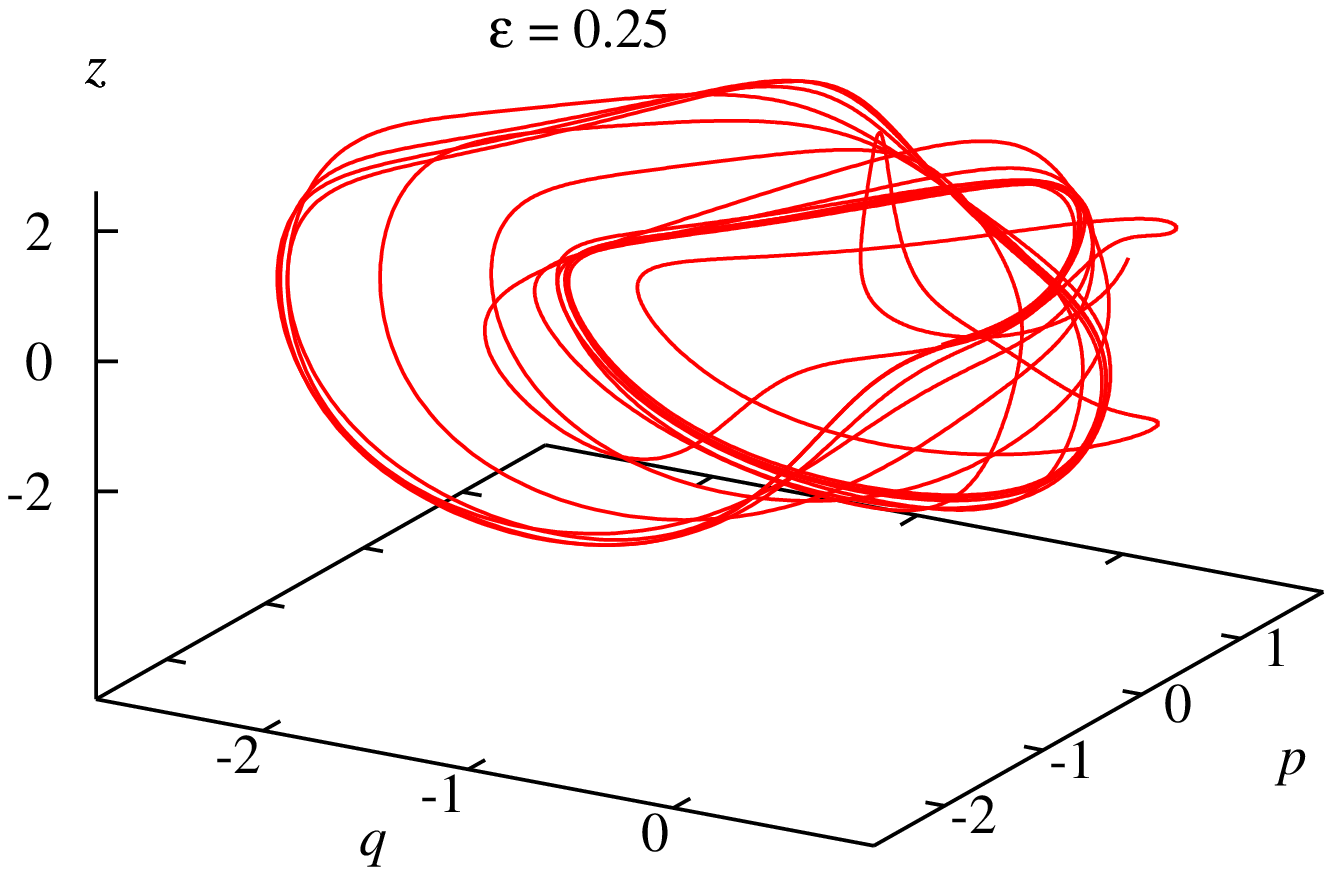}
\vspace{-1.0cm}
\caption{(Color online)  Projection of a short chaotic trajectory for $\varepsilon = 0.10 $ (top)
and $\varepsilon = 0.25 $ (bottom) onto the $qpz$-subspace.} 
\label{figure_3}
\end{figure}

\begin{figure}[ht]
\center
\includegraphics[width=0.5\textwidth]{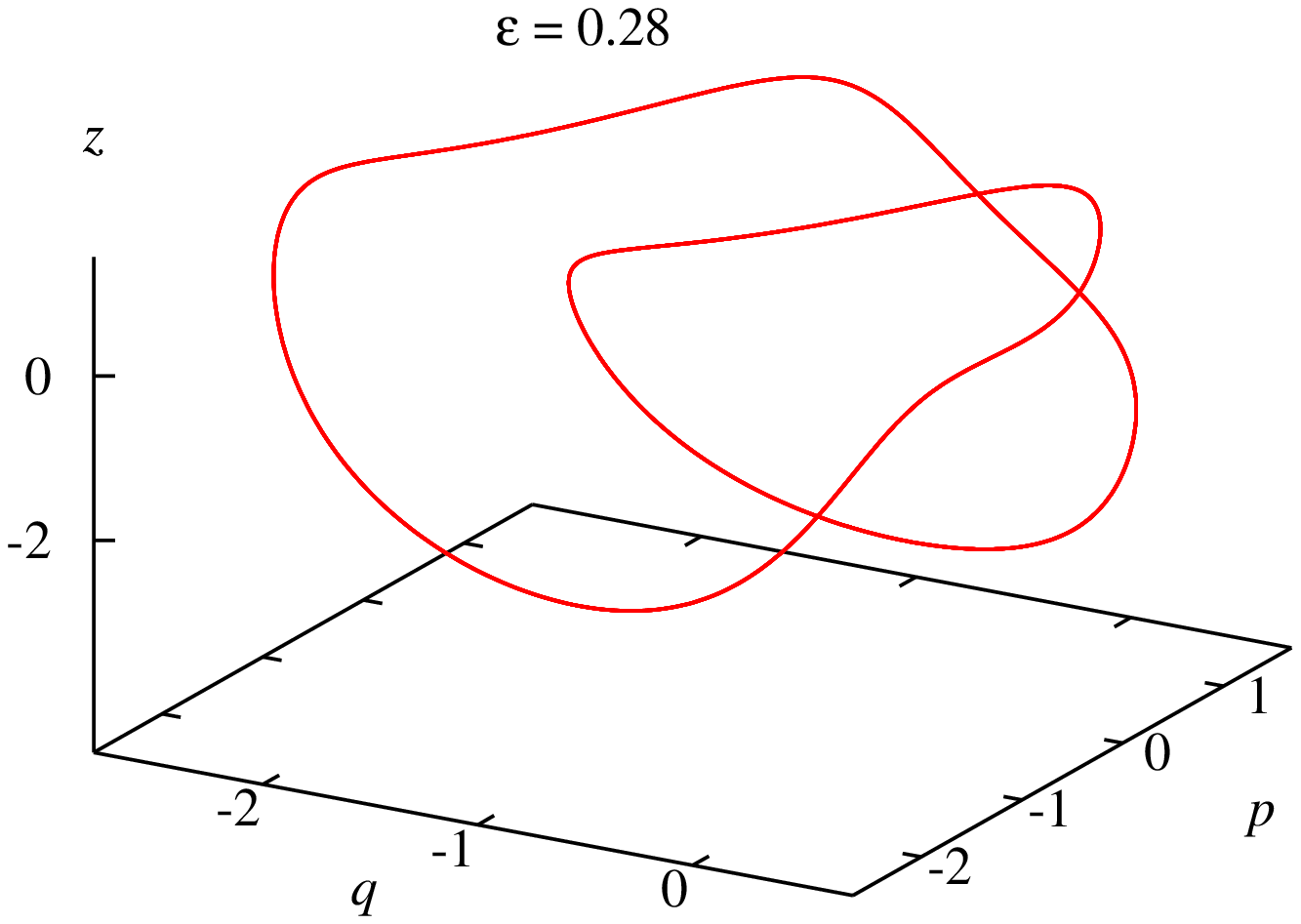}\\
\vspace{-1.0cm}
\includegraphics[width=0.5\textwidth]{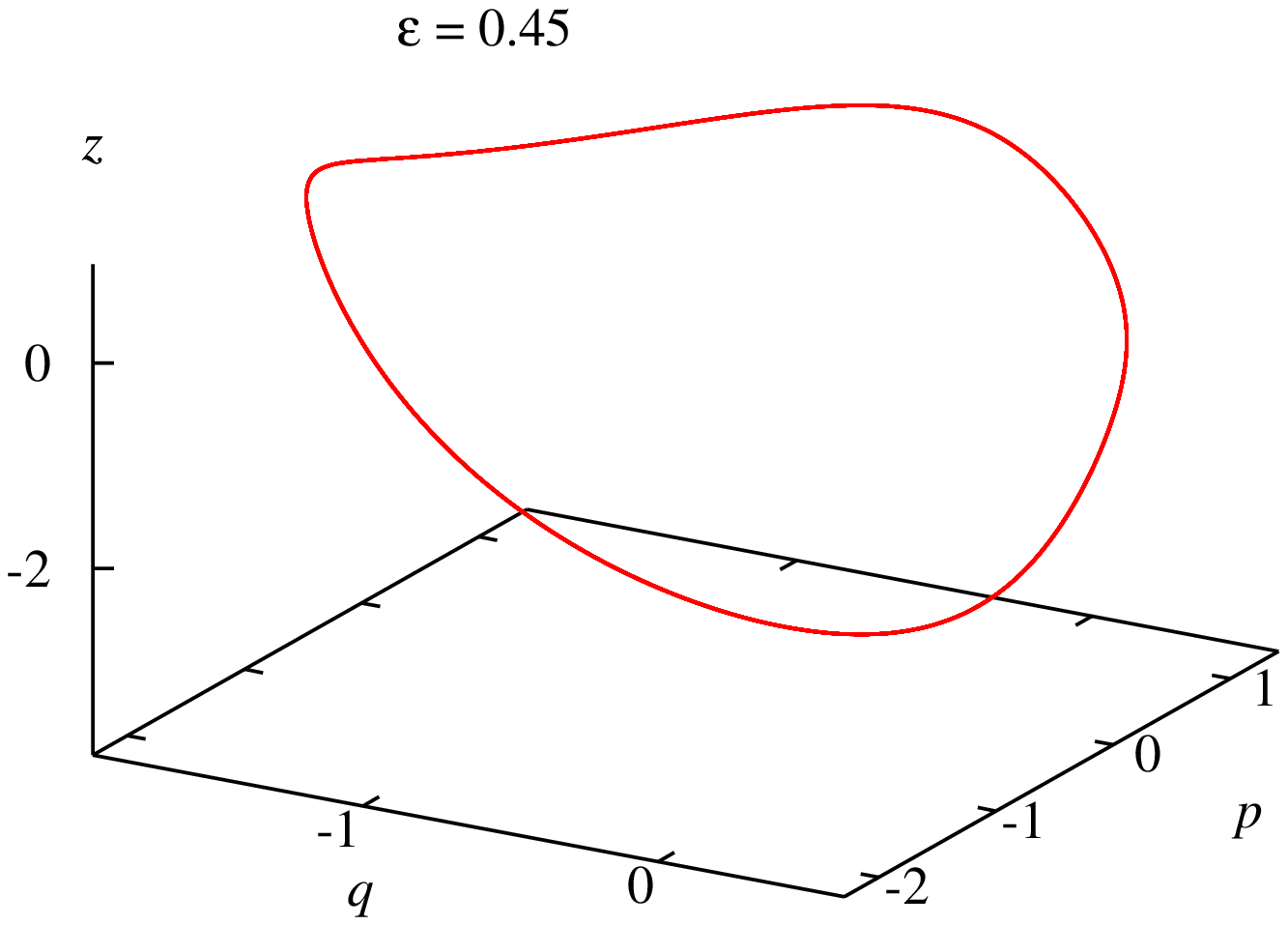}
\vspace{-1.0cm}
\caption{(Color online)  Projection of the limit cycle for $\varepsilon = 0.28 $ (top)
and $\varepsilon = 0.45 $ (bottom) onto the $qpz$-subspace.} 
\label{figure_4}
\end{figure}


\subsection{Local Lyapunov exponents}
\label{local_exponents}

 \begin{figure}[h]
\center
\includegraphics[width=0.5\textwidth]{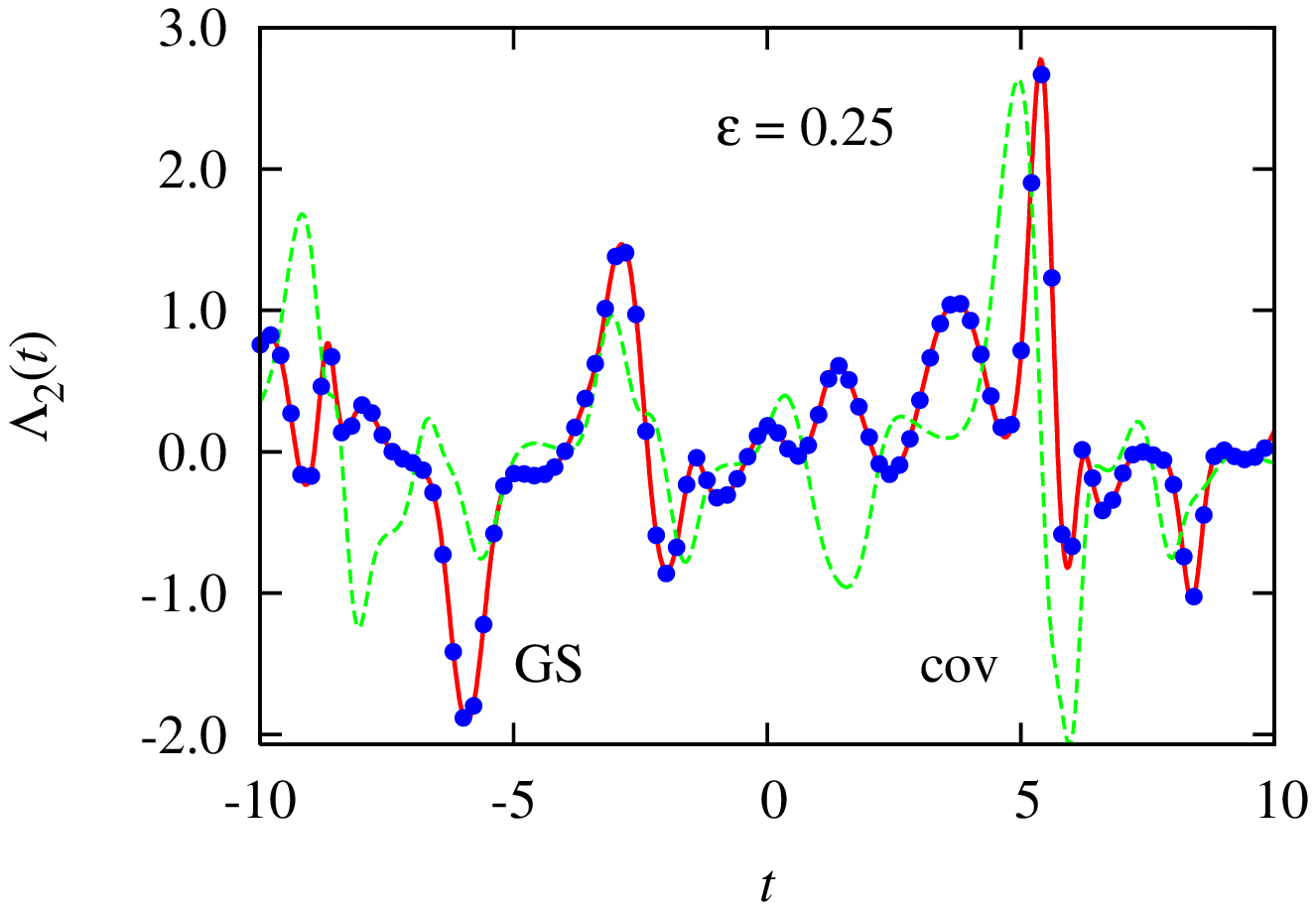} \\
\vspace{-5mm}
\caption{(Color online) 
Time-dependence of the local Lyapunov exponents $\Lambda_{2}(t)$ in the forward
direction of time for the doubly-thermostated oscillator with $\epsilon = 0.25$.
The  smooth red curve denotes
Gram-Schmidt exponents, directly obtained with a re-orthonormalization procedure, the blue 
points are computed with Eq. (\ref{inst_lya}), using the covariant time-dependent 
exponents $\Lambda_{\ell}^{\mbox{cov}}$  (also shown as green dashed line) and the angles 
$\beta_{\ell \ell}$ as input.  For clarity, only every 20th point is depicted.   }
\label{figure_5}
\end{figure}
In Fig. {\ref{figure_5} we apply Eq. (\ref{inst_lya}) to the
 doubly-ther\-mo\-stated oscillator in a stationary chaotic state, $\varepsilon = 0.25$.
For $\ell = 1$ the respective time-dependent exponents are identical and are not shown. 
The case $\ell =2$ is treated in the figure. The dashed green line denotes the covariant
local exponent, the smooth red line is for the local GS-exponents, which is directly obtained from the 
simulation invoking Gram-Schmidt re-orthonormalization. The time interval $\tau$ is $0.01$. 
The blue points  for $\Lambda_{\ell}^{\mbox{GS}}(t)$ are
computed with Eq. (\ref{inst_lya}), where the covariant exponent $\Lambda_{\ell}^{\mbox{cov}}(t)$   
and the angle $\beta_{\ell \ell}(t)$ are taken from the simulation. The agreement is convincing.
Similar results are also obtained for $\ell = 3$ and $4$ (not shown).

\begin{figure}[ht]
\center
\includegraphics[width=0.5\textwidth]{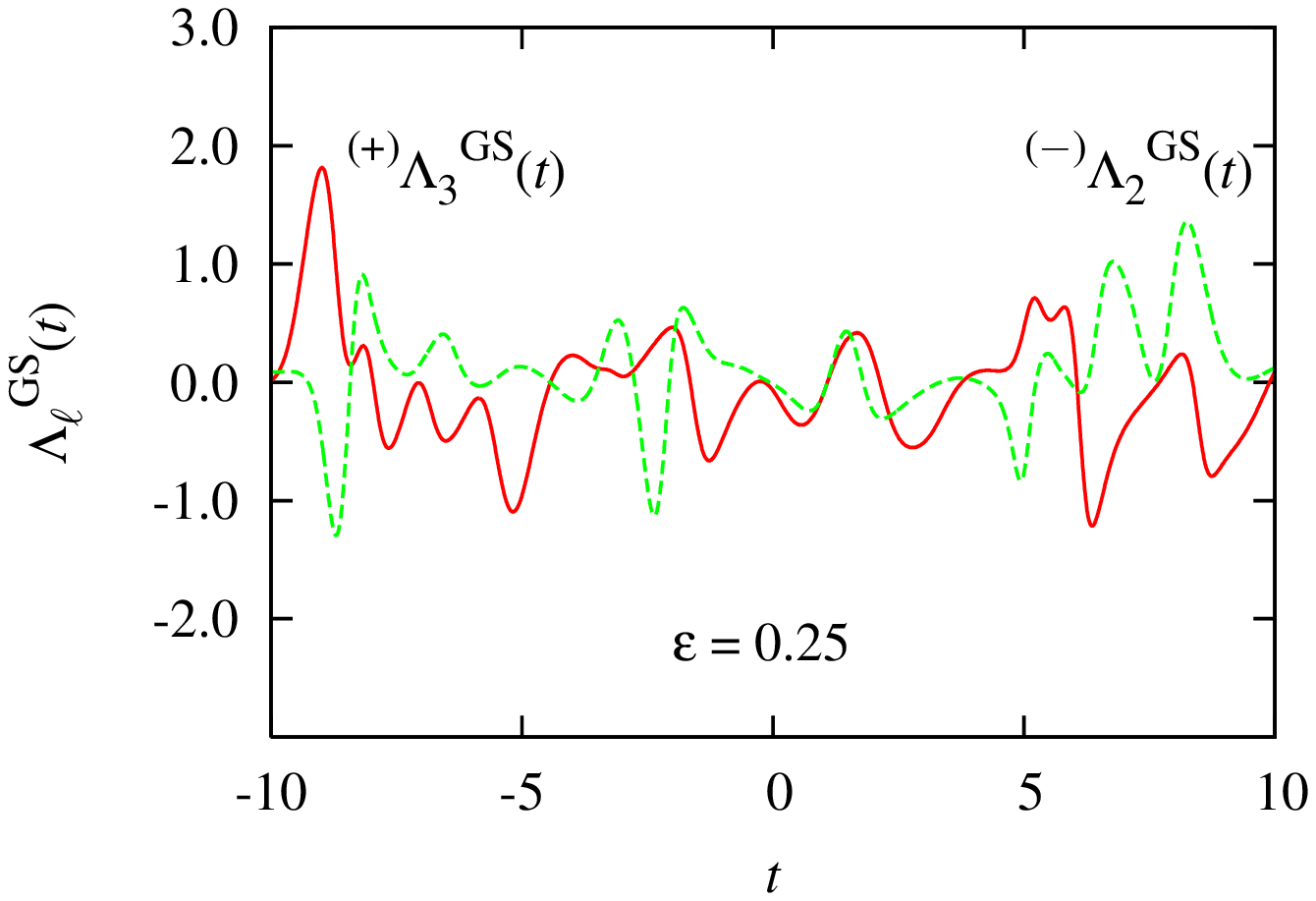}\\
\vspace{-5mm}
\includegraphics[width=0.5\textwidth]{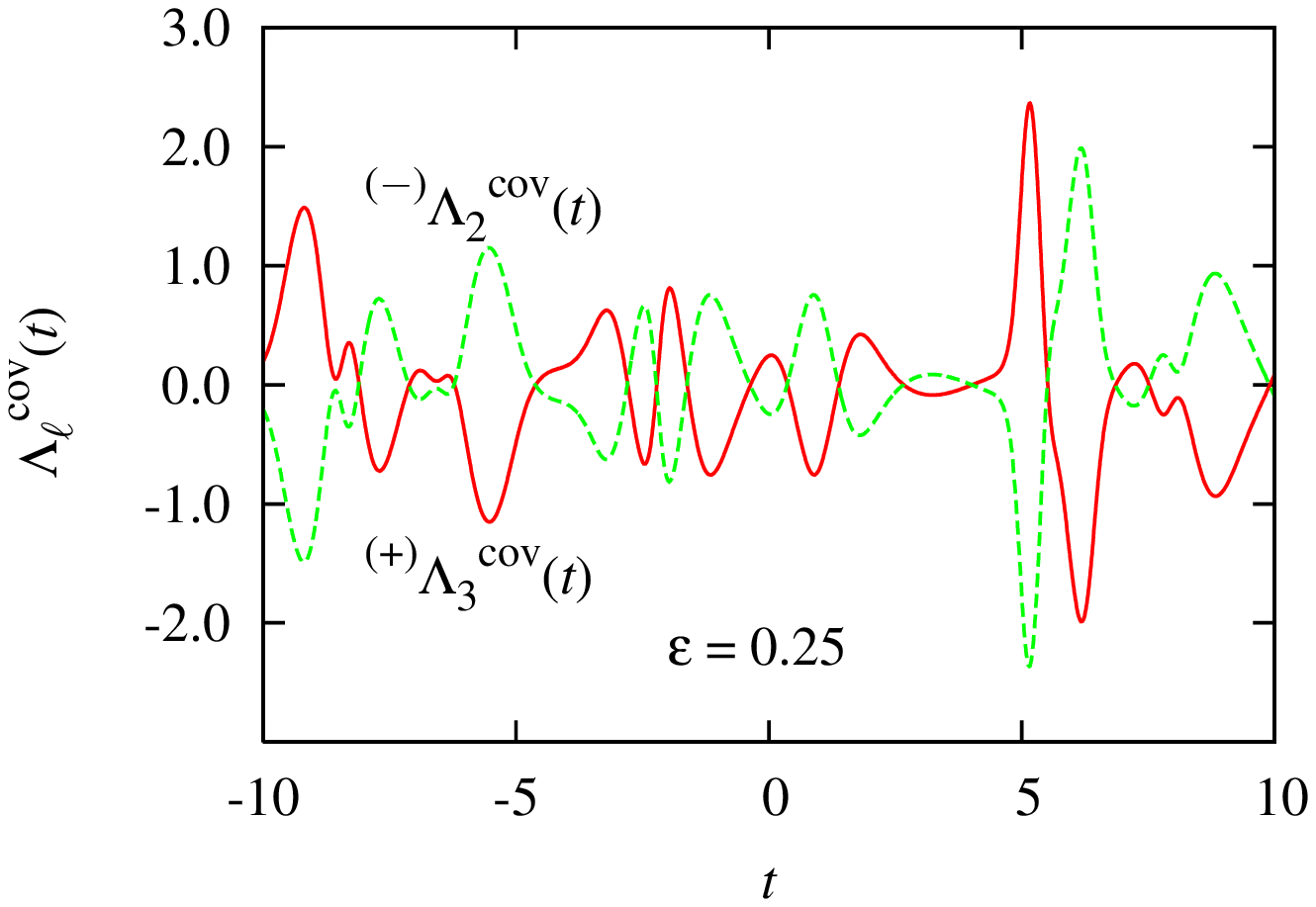}\\
\vspace{-0.5cm}
\caption{(Color online) Doubly-thermostated oscillator for $\epsilon = 0.25$.
Top panel: The Gram-Schmidt local Lyapunov exponents  do not display
 time-reversal symmetry.  Bottom panel: Display of time-reversal symmetry by the
 covariant local exponents,
$^{(+)}\Lambda_{\ell}^{\mbox{cov}} = - ^{(-)}\Lambda_{D+1 - \ell}^{\mbox{cov}}$ for $\ell = 2$,
 Analogous curves are obtained for the other $\ell$, but are not shown.}
\label{figure_6}
\end{figure}
In the bottom  panel of Fig. \ref{figure_6} we demonstrate, for $\ell = 2$,  the general 
time-reversal symmetry for the local (time dependent) covariant exponents (see Eq. (\ref{local_symmetry})  
which also gives rise to the symmetry of the global (time-averaged) exponents already 
encountered in Eq. ({\ref{global_exponent}). For $\ell = 1$ the symmetry is also fully 
obeyed but not shown.

As emphasized already in Eq. (\ref{not_equal}), the local Gram-Schmidt exponents generally 
do not have this symmetry. This is explicitly shown  in the
top panel of Fig. \ref{figure_6}. See also Ref. \cite{HPH01}, where the same observation was
made. Only the subspaces in Eq. (\ref{subspaces}) spanned by consecutive Gram-Schmidt vectors have a simple dynamical interpretation,  but not the GS-vectors themselves.
The orthonormal GS-vectors are oriented such that for the tangent
space, tangent to the phase flow at the phase point ${\vec \Gamma}(t)$, 
the subspaces ${^{(-)}\vec g}^{1}(t) \oplus \cdots \oplus ^{(-)}{\vec g}^{\ell}$, with 
$\ell \in \{1,\cdots ,D\}$, are the most unstable subspaces of dimension $\ell$ going from time $t$ to
$-\infty$ (i.e. the most stable subspaces of dimension $\ell$ in the future).
Although time reversal converts a most stable subspace of dimension $\ell$
into the most unstable subspace with the same dimension, and {\em vice versa}, there is no
obvious correlation of the instantaneous Lyapunov exponents 
$ ^{(-)}\Lambda_{\ell}^{\mbox{GS}}(t)$  and $^{(+)}\Lambda_{D+1 -\ell}^{\mbox{GS}}(t)$ for  $ \ell = 1,\cdots,D$.

\begin{figure}[htb]
\center
\includegraphics[width=0.5\textwidth]{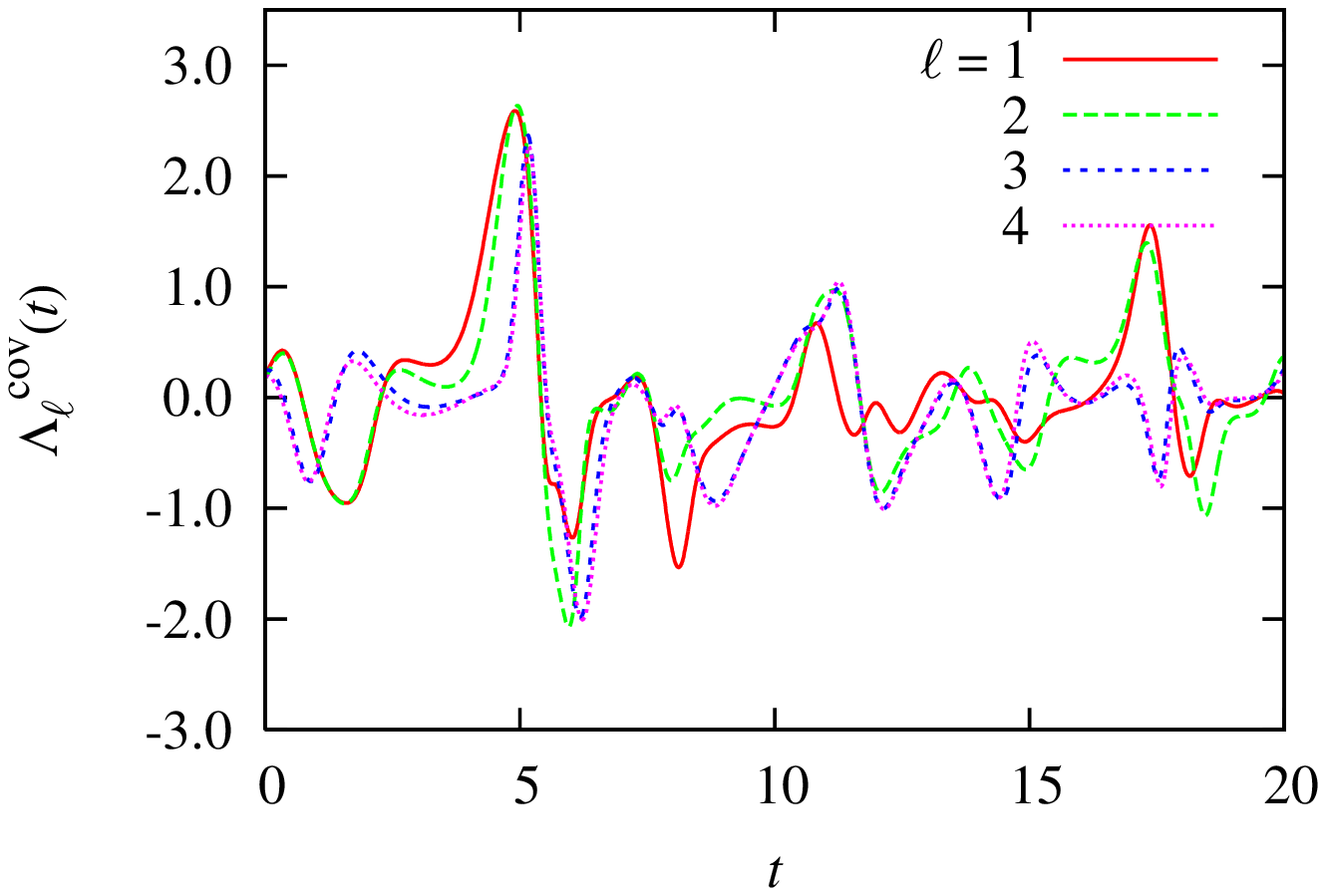}\\
\vspace{-0.5cm}
\caption{(Color online) Doubly-thermostated oscillator with $\epsilon = 0.25$: Time dependence of 
all four  local covariant Lyapunov exponents. }
\label{figure_7}
\end{figure}
It is interesting to follow the time dependence of the covariant local exponents, or more correctly expressed, 
their variation for consecutive state points along the phase space trajectory (see Fig. 
\ref{figure_7}). One observes that the order of the exponents fluctuates and may even be totally reversed
with $\Lambda_1^{\mbox{cov}}(t)$ being most negative and $\Lambda_4^{\mbox{cov}}(t)$ most
positive. Also the dimension of the stable and unstable manifolds changes along the trajectory.
This indicates that the system is far from being hyperbolic. We address this point more closely 
in the following subsection.


\subsection{Hyperbolicity}


\begin{figure}[ht]
\center
\includegraphics[width=0.5\textwidth]{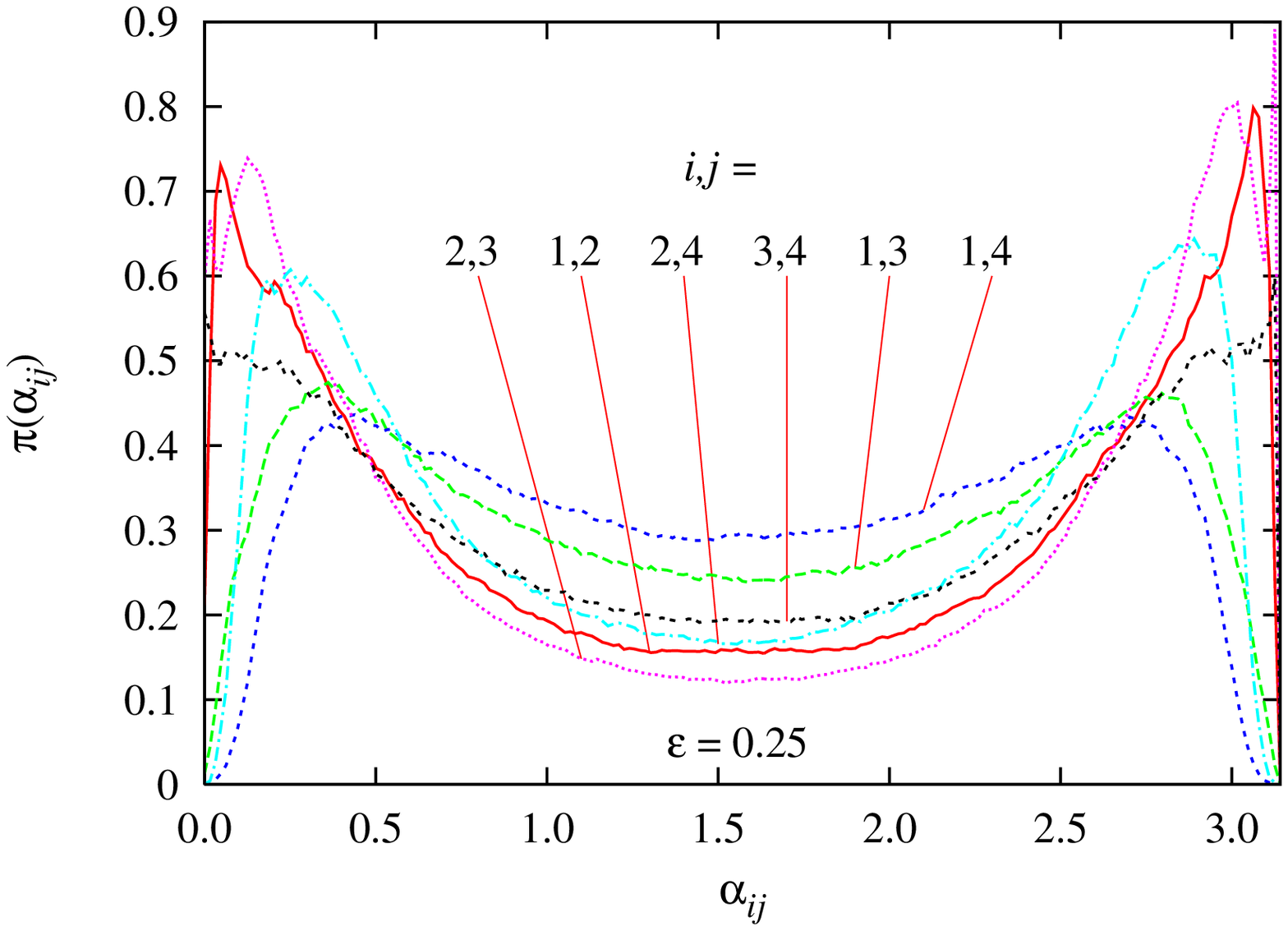}\\
\includegraphics[width=0.5\textwidth]{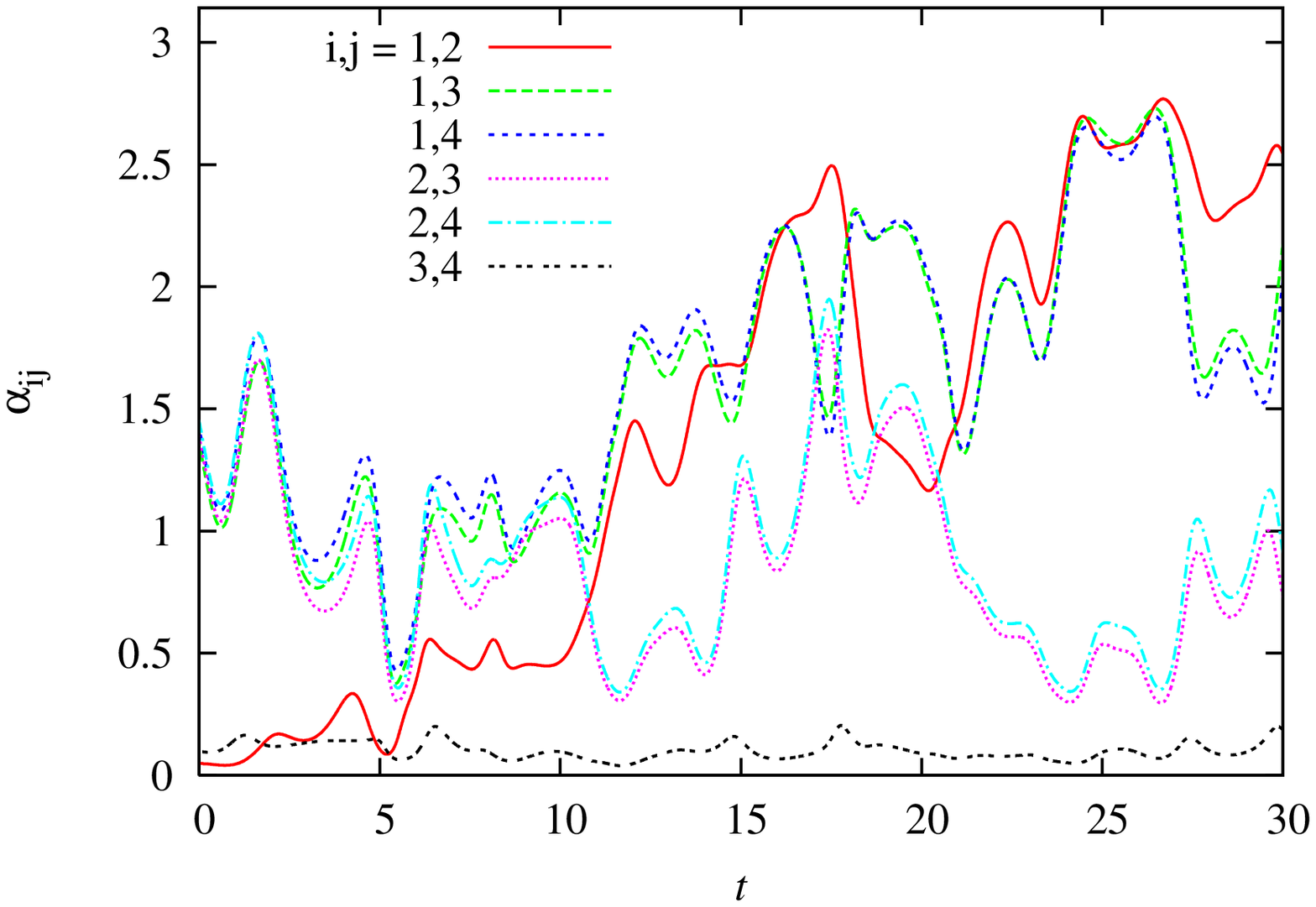}\\
\vspace{-0.5cm}
\caption{(Color online) Doubly-thermostated oscillator with $\epsilon = 0.25$. Top panel: 
Probability distributions for the angles $\alpha_{i j}$ between covariant
vector pairs specified by the labels.  Bottom panel:  Time evolution of the angles
 $\alpha_{i j}$ between the covariant vectors ${\vec v}^i$ and ${\vec v}^j$. }
\label{figure_8}
\end{figure}
We infer from Eq. (\ref{inst_lya}) that the difference between the local covariant and Gram-Schmidt 
exponents stems from the fact that the angle between the respective vectors deviates 
significantly from zero and varies with time. But also the angles between covariant vectors,
$ \alpha_{i j}(t) \equiv \arccos[({\vec v}^i \cdot {\vec v}^j ) / | {\vec v}^i | | {\vec v}^j |    ]$ significantly change
with time. This is demonstrated in the bottom panel of Fig. \ref{figure_8} for the same 
nonequilibrium state ($\varepsilon = 0.25$) of the doubly-thermostated oscillator  
discussed previously. There is an intermittent tendency of any two pairs of vectors to get parallel
or antiparallel to each other. The probability distributions for these angles, $\pi(\alpha_{i j})$, are 
shown in the  top panel of Fig. \ref{figure_8} and confirm this observation. 
Although the angles $\alpha$  do not become strictly zero -- the vectors could not 
separate anymore after such an event, which is not observed --
the large probability for angles  close to zero or $\pi$ is noticeable. 
As was mentioned before, the associated local covariant exponents
are out of order for most of the time as in Fig. \ref{figure_7}.  If $P_i$ denotes the
probability for $\Lambda_i^{\mbox{cov}}$ to be out of order  with respect to any of 
the other exponents, one finds for the doubly-thermostated oscillator $(\varepsilon = 0.25)$ 
$\{ P_1, \cdots, P_4 \}  = \{ 0.650, 0.813, 0.840, 0.645 \}$. This clearly demonstrates
the strong entanglement between the covariant vectors.  If the local exponents are
time averaged along the trajectory for  time intervals $\Delta$,  the analogous
probabilities $\overline{P}_i^{\Delta}$ for the time-averaged exponents 
$\overline{\Lambda_i^{\mbox{cov}}}^{\Delta}$ scale according to 
$\overline{P}^{\Delta} \propto \Delta^{-\gamma_i}$  with $\gamma > 0$ for large-enough  $\Delta$.
This shows that the domination of the Oseledec splitting is violated for finite times. 

Such a behavior is  in contrast to the 
covariant dynamics of hard-disk systems, for which the covariant vectors tend to avoid
becoming parallel or antiparallel \cite{BP10}. Thus, whereas the hard-disk  system is hyperbolic,
the doubly-thermostated oscillator is not.


\subsection{Singularities of the local Lyapunov exponents}

   In the direction of the flow, the local Lyapunov exponents 
clearly are smooth functions of the time and, hence, of the phase-space position along the
trajectory, see Fig. \ref{figure_5}. But transverse to the flow 
this need not be the case. Indeed, for the periodic Lorentz gas it was noted by Gaspard \cite{Gaspard1,Gaspard2} that the local stretching factors are discontinuous transverse 
to the flow. Since this model involves hard elastic collisions of point particles with  space-fixed scatterers,
the observed discontinuity might still be thought to be a consequence of the discontinuous nature of
the flow. However, Dellago and Hoover showed \cite{Dellago_Hoover_2000} that this is not the case.
They found a discontinuous local exponent $\Lambda_1^{\mbox{GS}}$ along a path transverse
to the flow even for a time-continuous Hamiltonian system, a chaotic pendulum on a spring.
Of course, their result also applies to    $\Lambda_1^{\mbox{cov}}$ for that model. Here we 
provide evidence for the doubly-thermostated oscillator in equilibrium ($\varepsilon = 0$)
that all local covariant exponents are discontinuous along directions transverse to the flow.
 
     For this simulation we slightly modify the  protocol of Section \ref{numerics}. \\ 
{\bf Phase 0}: Starting at a phase point ${\vec \Gamma}_s$ at time zero, 
the reference trajectory is followed backward in time to $-t_{\omega} = -60,000$ and is 
periodically stored for intervals $\tau$ along the way. \\
{\bf Phase 1}: The next  phase is identical to phase 1 of Section
\ref{numerics}  with one essential difference: For  $-t_{\omega} \le t \le 0$, the previously-stored reference trajectory is now used in the forward direction of time for the computation of the 
Gram-Schmidt vectors, which assures that the trajectory precisely arrives  at 
${\vec \Gamma}_s$ at time zero in spite of the inherent Lyapunov instability.  For
$0 \le t \le t_{\omega}$  the simulation proceeds as in phase 1 of Section \ref{numerics}.\\ 
{\bf Phase 2}: This is identical to  phase 2 of  Section \ref{numerics} and  provides us with
the covariant vectors and the respective local exponents in the interval 
$-t_0 < t < t_0$ and at the time $t  = 0$ in particular, when the state coincides with the 
selected phase point ${\vec \Gamma}_s$.

 The whole  procedure is repeated for starting points ${\vec \Gamma}_s \equiv {\vec \Gamma}_0 + s \times ( 0,0,1,0 ) $ on a straight 
line parallel to the $z$-axis, which is parametrized by $s$.  This line is transversal to the flow, as may be expected from Fig. \ref{figure_5}.

      As an example, we plot in the top panel of Fig. \ref{figure_9} the local covariant exponent
$\Lambda_4^{\mbox{cov}}(t) $ as a function of time  for 500 
initial points ${\vec \Gamma}_s$ separated by  $\Delta s = 1 \times 10^{-4}$. 
It should be noted that the scale on the $t$ axis, converted into distances
along the trajectory, is about 200 times coarser than that on the $s$ axis.
The red line for $t = 0$  connects the local exponents for the selected initial states 
${\vec \Gamma}_s$. This curve for $\Lambda_4^{\mbox{cov}}(t) $ is also reproduced in  
the bottom panel of Fig.  \ref{figure_9} together with an analogous result for 
$\Lambda_1^{\mbox{cov}}(t) $.  Both curves  exhibit singularities on many scales
showing singular  fractal character. There are no obvious correlations between the two curves.
The singularities are due to bifurcations in the past history of the trajectory. In view of
Fig. \ref{figure_5}, such a bifurcation may be visualized, for example, by 
a transition of the trajectory from the neighborhood of an unstable periodic orbit to 
the neighborhood of another with a different number of loops. 
\begin{figure}[ht]
\center
\vspace{-0.5cm}
\includegraphics[width=0.55\textwidth]{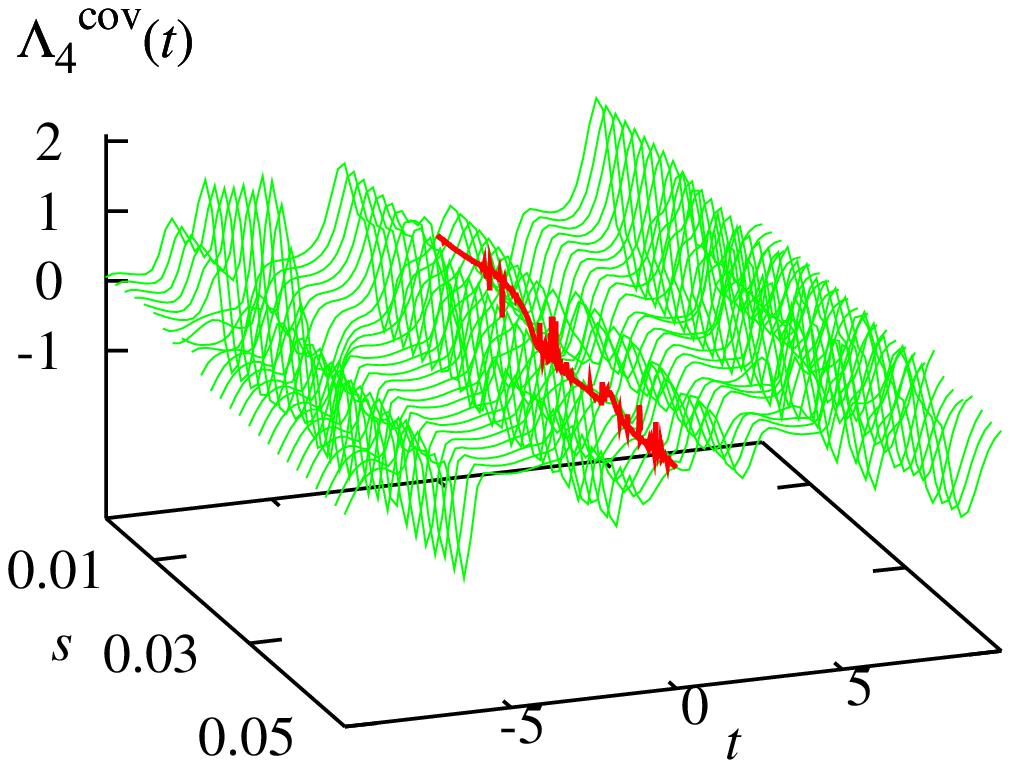}\\
\vspace{-8mm}
\includegraphics[width=0.5\textwidth]{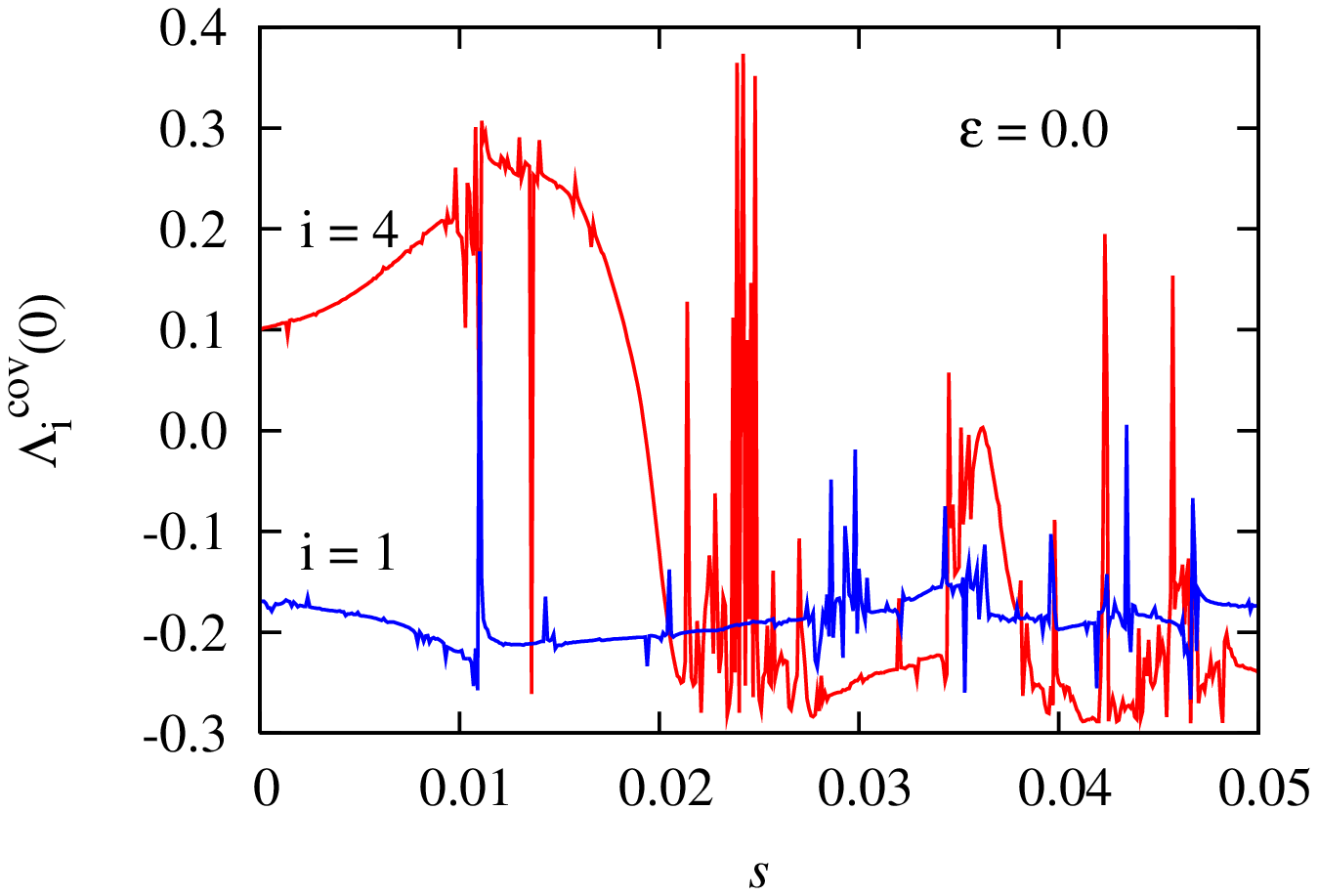}\\
\vspace{-0.5cm}
\caption{ (Color online) Covariant local Lyapunov exponent, $\Lambda_4^{\mbox{cov}}$, for the  
doubly-thermostated oscillator in equilibrium. Top panel: Time dependence  along trajectories,
which, for $t = 0$, are at the specified initial points  ${\vec \Gamma}_s$ introduced in the main text.
These phase points lie on a straight line transverse to the flow, and  $s$ specifies the precise position.
The variation of the local exponent along this straight line is shown as a red  curve.
Bottom panel: The red curve is a magnified view of the red line of the upper panel,
representing the variation of $\Lambda_4^{\mbox{cov}}$ along a straight line
transverse to the flow in the phase space. The blue line is an analogous curve for 
$\Lambda_1^{\mbox{cov}}$.  }
\label{figure_9}
\end{figure}

   One may raise the question (as has been done by one of the referees), how reliable the
curves in Fig. \ref{figure_9} are in view of the chaotic nature of the flow
and problems of shadowing due to the finite computational accuracy. An increase of the
Runge-Kutta integration time step $dt$ by a factor of four has no noticeable effect (less than 0.1\%) 
in Fig. \ref{figure_9}, which also proved completely insensitive to a reduction of the
relaxation time $t_{\omega}$ of the algorithm by a factor of two and of an increase of  the time $\tau$
between successive re-orthonormalization steps by the same factor.  This robustness,
however, does not apply to  the local exponents  $\Lambda_2^{\mbox{cov}}$ and 
$\Lambda_3^{\mbox{cov}}$ (not shown), which belong to the two-dimensional central manifold 
for this equilibrium system. The respective covariant vectors span this subspace, but their
precise orientations and their local exponents are affected by details of the algorithm and 
do not have direct physical significance.

   For nonequilibrium stationary states the singular character of the local exponents in transverse 
directions is expected to be even  more pronounced, since even the phase-space
probability density becomes a multifractal object \cite{HHP87,PH88}. For the covariant exponent this cannot
be shown with the present algorithm. The reason is that during the time-reversed 
simulation in phase 0,  the phase volumes collapse yielding negative Lyapunov exponent sums.
Since in phase  1 this trajectory is followed in the opposite direction, the respective
phase volumes {\em expand} providing a positive sum of Lyapunov exponents, but only up
to time zero. For positive times the reference trajectory is calculated anew from the motion
equations, again yielding {\em contracting} phase volumes. Thus, the character of the 
flow  changes at $t = 0$ and the Gram-Schmidt vectors at first are non-relaxed and point into wrong 
directions for positive times. Since these vectors are required for the computation 
of covariant vectors at and near zero time,
the algorithm cannot be used to obtain the covariant vectors and respective local exponents
at a pre-determined point ${\bf \Gamma}_s$ in phase space. For equilibrium states this
restriction does not apply and the local exponents may be computed for
pre-specified phase-space points.
  

\section{Local Lyapunov exponents for symplectic systems}
\label{sym_systems}

So far we have omitted to mention that we use a particular metric in phase space. Whereas the
global exponents are independent of the metric, the local exponents, covariant or Gram-Schmidt, clearly 
are not. We demonstrate this with the most simple symplectic example, a scaled harmonic oscillator 
\cite{HHP90,HHG08} with Hamiltonian
$$
   H_s (p,q)= \frac{1}{2}\left[ \left( \frac{p}{s}    \right)^2   + (s q)^2  \right]
 $$  
 and equations of motion 
 \begin{equation}   
 \dot{q} =  s^{-2} p \;, \;\;\;  \dot{p} = -s^2 q,  
 \label{eom}
 \end{equation}
 where $s$ is a scaling parameter.
 For the `natural' scaling, $s = 1$, the global and local exponents vanish. But for
\begin{figure}[htb]
\center
\includegraphics[width=0.5\textwidth]{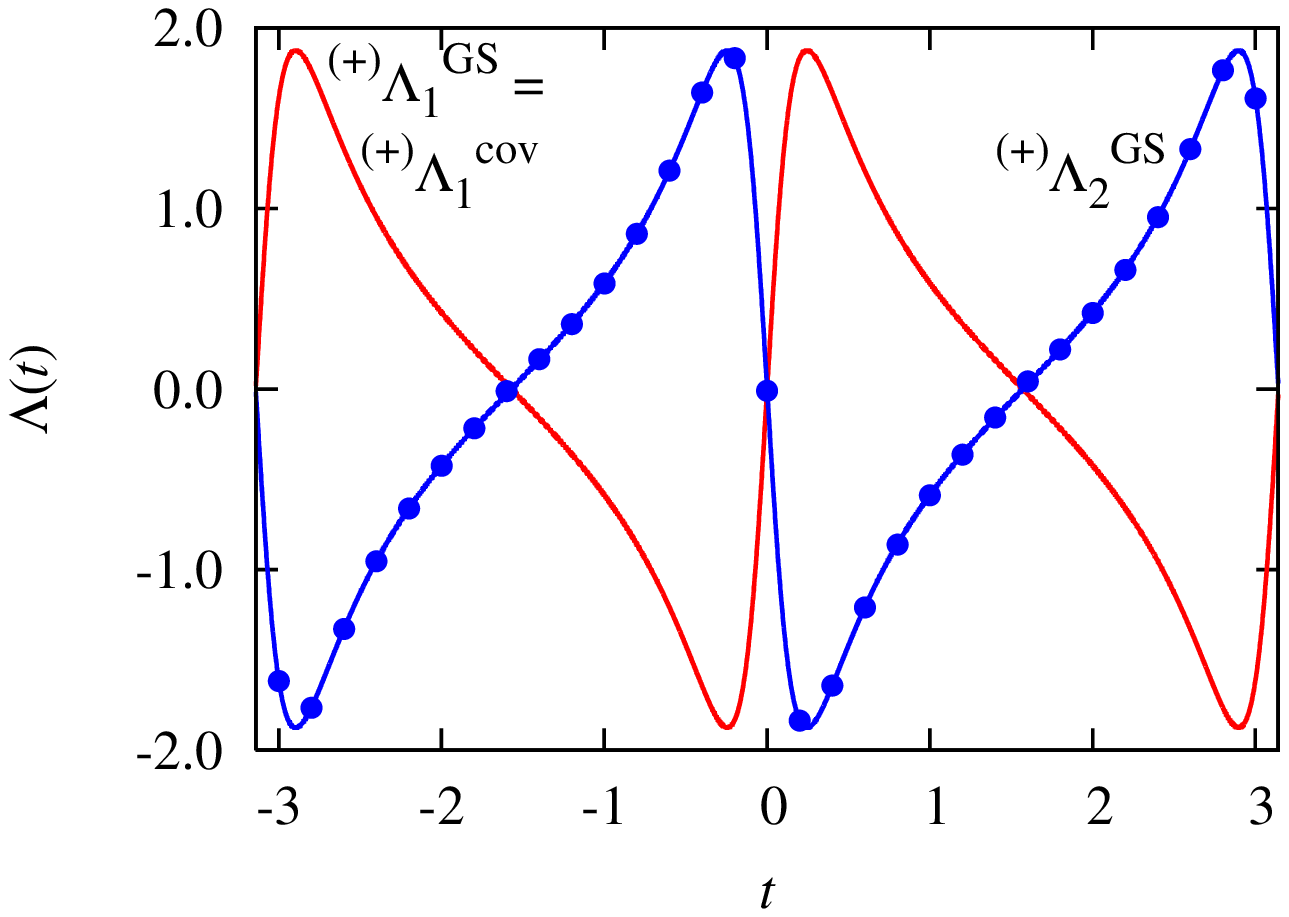}\\
\vspace{-5mm}
\includegraphics[width=0.5\textwidth]{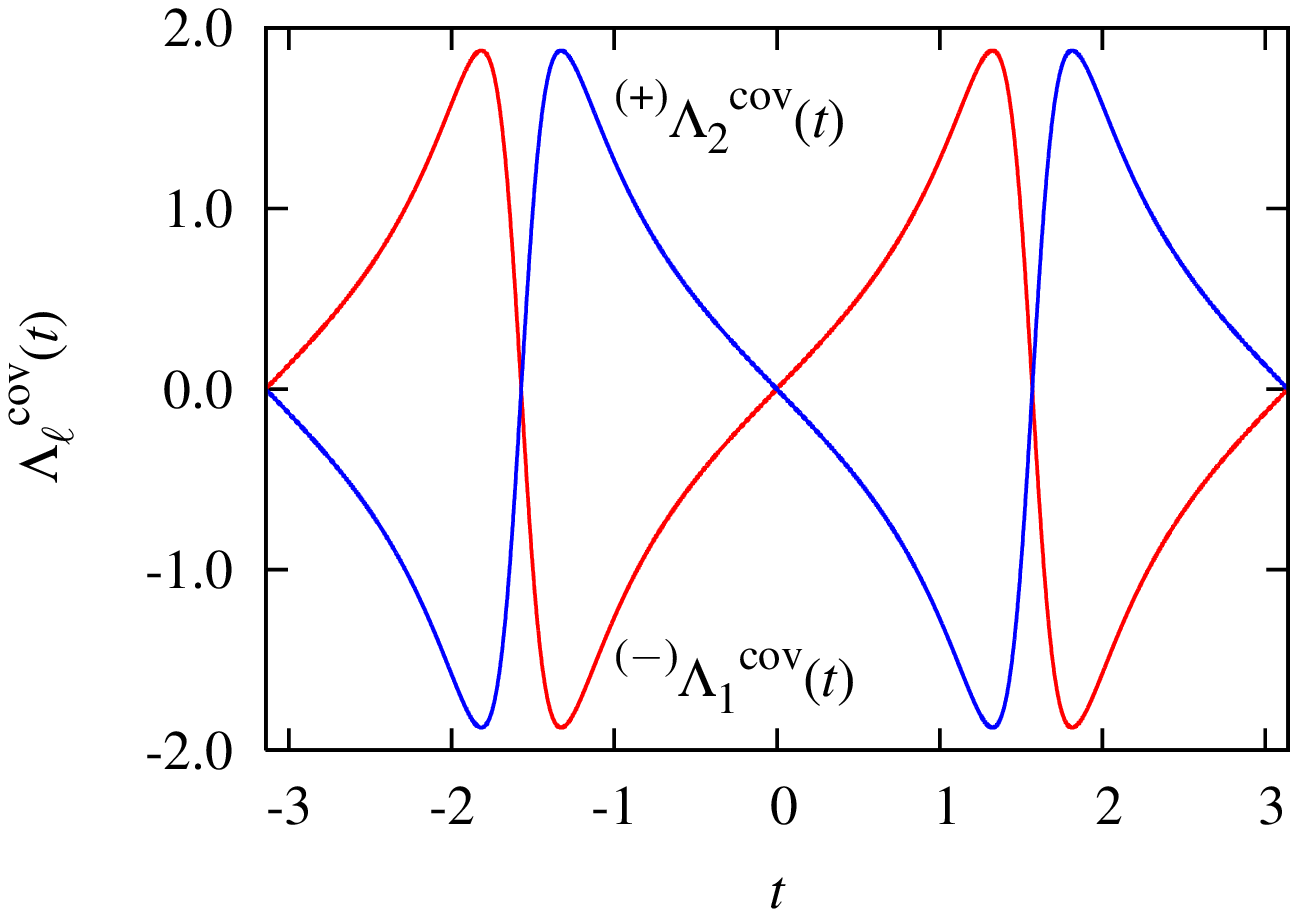}\\
\vspace{-0.5cm}
\caption{(Color online) Scaled harmonic oscillator with $s = 2$: Time dependence of 
the local Lyapunov exponents. 
Top panel: The smooth red lines are local GS exponents
for $\ell = 1$ and $2$ as indicated by the labels. The results from direct simulation and
from theory are undistinguishable on this scale.
The points indicate the reconstruction  of $^{(+)}\Lambda_2^{\mbox{GS}}$  with the help of Eq. (\ref{inst_lya}),
where $^{(+)}\Lambda_2^{\mbox{cov}}$ is shown in the lower panel.
Note that the GS and covariant exponents for $\ell = 1$ are identical.
Bottom panel: Demonstration of the time-reversal invariance for the covariant exponents.   }
\label{figure_10}
\end{figure}
the scaled case, $s \ne 1$,  the local Lyapunov exponents do not. They depend on the metric 
and, for that matter, on the choice of the coordinate system, be it Cartesian or polar.

Let us look at this model in a little more detail, since the dynamical matrix 
$$\cal J  =   \begin{pmatrix} 0 & s^{-2} \\  -s^2 & 0\end{pmatrix}$$  
for this linear model does not depend on the phase point and allows for a complete analytical solution
\cite{HHG08} for the tangent vector dynamics. Still, 
the model  is a bit peculiar, since there is no global exponential
ordering of tangent vectors familiar from the Gram-Schmidt algorithm,
and the considerations of Sec. \ref{numerics} lose their meaning. 
Any (unit) vector, with  arbitrary initial condition (phase), which is  a solution of Eq. (\ref{l1}), may be taken as the 
first Gram-Schmidt vector (or covariant vector for that matter). We  fix this arbitrary phase by requiring that 
${\vec g}^1$ coincides with the normalized phase-space velocity, which is associated with a
vanishing exponent and is a solution of Eq. (\ref{l1}), as may be explicitly shown. Thus,
 \begin{equation} 
   {\vec g}^1 \equiv  \begin{pmatrix} \delta q_1 \\ \delta p_1 \end{pmatrix}
      = \left[  \left(\frac{p}{s^2}\right)^2 + (s^2 q)^2         \right]^{-1/2} \begin{pmatrix} p/s^2  \\ -s^2 q \end{pmatrix},
\label{g1}
\end{equation}      
where we denote the perturbation components of 
${\vec g}^1$ by $  \delta q_1$ and $\delta p_1)$.
From the constrained motion equation (\ref{l1}) one obtains
\begin{eqnarray}
\dot{\delta q}_1 & = & s^{-2} \delta p_1   -  R_{11} \delta q_1 \label{d1}\\
\dot{\delta p}_1 & = & -s^2     \delta q_1  - R_{11} \delta p_1. \label{d2} \\
\nonumber
\end{eqnarray}  
Since $ {\vec g}^1$ is constrained to unit length, $ \delta q_1 \dot{\delta q}_1 + \delta p_1 \dot{\delta p}_1 = 0$,
and noting that $\Lambda_1^{\mbox{GS}} \equiv R_{11}$, we obtain from these equations
\begin{equation}
         \Lambda_1^{\mbox{GS}} = \left(s^{-2} - s^2\right) \delta q_1 \delta p_1.
\label{GS1}
\end{equation}
Insertion of $\delta q_1$ and $\delta p_1$ from Eq. (\ref{g1}) yields
an analytic  expression for the local Gram-Schmidt exponent $\Lambda_1^{\mbox{GS}}$
as a function of the phase-space point $(q,p)$. The second Gram-Schmidt vector is perpendicular to 
the first, and its associated GS exponents immediately follow from the conservation of phase-space volume:
$$
         \Lambda_2^{\mbox{GS}}(q,p) =  -\Lambda_1^{\mbox{GS}}(q,p).
$$
In the top panel of Fig.  \ref{figure_10} the local Gram-Schmidt exponents for the scaled
harmonic oscillator for $s=2$ are shown as a function of time. The
 initial conditions $ q(0) = 0$ and $ p(0) = 1$  are used, for which the solution of Eq. (\ref{eom})
becomes
 $$
      q(t) = s^{-2} \sin t \;, \;\;\; p(t) = \cos t.
 $$     
Both computer simulation results and the theoretical expressions for  $\Lambda_1^{\mbox{GS}}$ and 
$\Lambda_2^{\mbox{GS}}$
are shown,  which agree so well that they  cannot be distinguished and appear only as 
a single smooth red line  in the figure.  

The upper and lower bounds of the local exponents   
are also easily obtained from Eq. (\ref{GS1}) \cite{HHG08},
$$ 
      \Lambda_{\mbox{min,max}} = \pm (s^{+2} - s^{-2}) / 2.
$$      
The extrema are attained whenever the  components $\delta q$ and $\delta p$ 
of the respective perturbation vectors contribute equally,  $\delta p = \pm \delta q$. For $s = 2$ 
we find $  \Lambda_{\mbox{min,max}} = \pm 1.875$ in full agreement with Fig. \ref{figure_10}.

The undetermined phases we encountered with the Gram-Schmidt vectors also
carry over to the covariant vectors. But since the latter are computed with the
help of the former, the choice of phase for ${\vec g}^1$ also fixes
that for ${\vec v}^1$ and    ${\vec v}^2$. In the lower panel of Fig. \ref{figure_10} 
the covariant exponent $\Lambda_2^{\mbox{cov}}(t)$  is shown as it is obtained from the simulation. 
If this function is used to reconstruct $\Lambda_2^{\mbox{GS}}(t)$ with the help of
Eq. (\ref{inst_lya}), the (blue) dots in the upper panel of Fig. \ref{figure_10}  are 
obtained, where $\cos(\beta_{22}) = {\vec g}^2 \cdot {\vec v}^2 $ is also taken from the simulation.
The agreement is very good. In the lower panel of Fig. \ref{figure_10} we also plot  $^{(-)}\Lambda_1^{\mbox{cov}}(t)$ for the 
time-reversed dynamics. The time-reversal symmetry of Eq. (\ref{local_symmetry}),
$$
 ^{(-)}\Lambda_{1}^{\mbox{cov}}(t)  = -^{(+)}\Lambda_{2}^{\mbox{cov}}(t) ,
$$
is nicely displayed.

As a slightly more involved example, we compute the four local Lyapunov
exponents for the symplectic H\'enon-Heiles system \cite{Henon} with Hamiltonian
\begin{equation}
 H = \frac{1}{2} (p_x^2 + p_y^2 ) + \frac{1}{2} ( x^2 + y^2 ) + x^2 y - \frac{1}{3} y^3.
 \label{henon}
\end{equation}
For an energy $H = 1/6$, the system is known to be chaotic
(with a Lyapunov spectrum $\{ 0.127_7, 0, 0, -0.127_7 \}$), where the trajectory visits 
most of the accessible phase space \cite{LL_Buch,Ramaswamy}. 
\begin{figure}[tb]
\center
\includegraphics[width=0.5\textwidth]{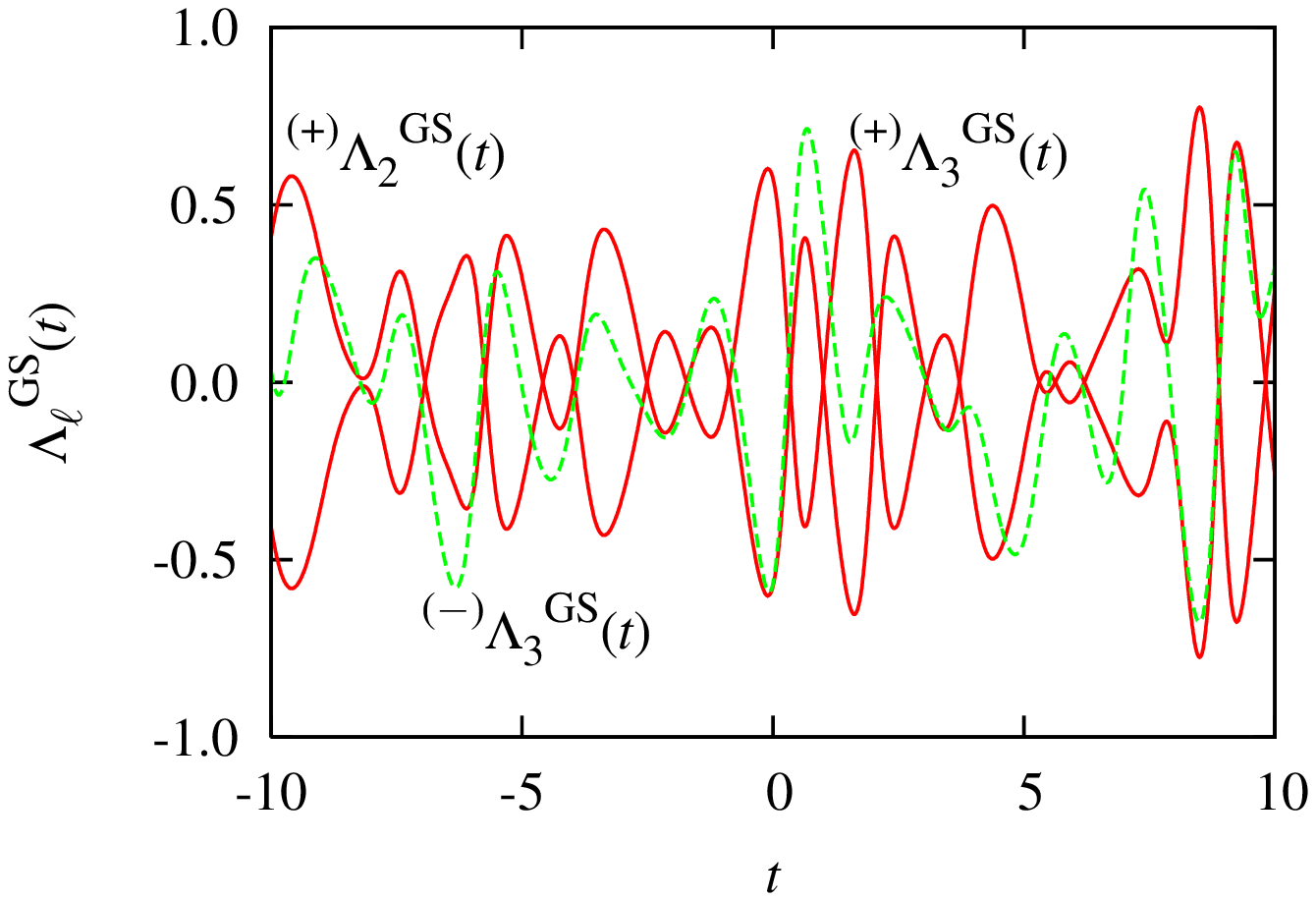}\\
\vspace{-5mm}
\includegraphics[width=0.5\textwidth]{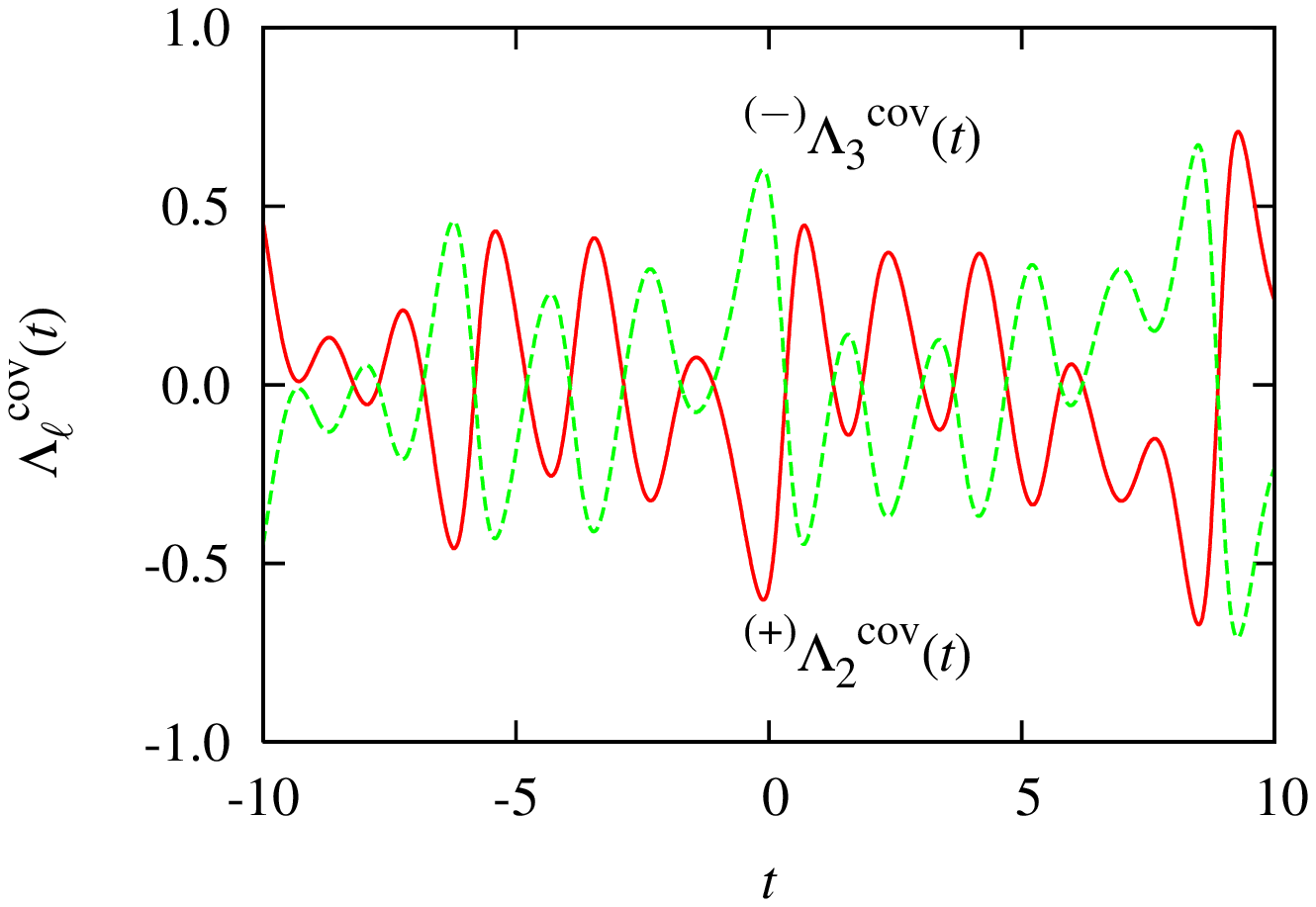}\\
\vspace{-0.5cm}
\caption{(Color online) Local Lyapunov exponents for the H\'enon-Heiles system
with  energy $H = 1/6$. Top panel: Illustration of the symplectic local pairing symmetry,
Eq. (\ref{gsfwd}), for the Gram-Schmidt exponents $^{(+)} \Lambda_2^{\mbox{GS}}$
and $^{(+)}\Lambda_3^{\mbox{GS}}$ (smooth red lines). Here, $D = 4$.
Also the inequality Eq. (\ref{not_equal}) applies ($\ell = 2$), as the dashed green
line for $^{(-)}\Lambda_3^{\mbox{GS}}$ certifies. Bottom panel: Verification of the 
time-reversal invariance property (\ref{local_symmetry}) for the covariant vectors specified.} 
\label{figure_11}
\end{figure}
Using the protocol of  Section \ref{numerics}, we compute the GS and covariant exponents 
and present some of the results in Fig. \ref{figure_11}. In the top panel the
symplectic local pairing symmetry of Eq. (\ref{gsfwd}) for $^{(+)}\Lambda_2^{\mbox{GS}}$
and $^{(+)}\Lambda_3^{\mbox{GS}}$ is shown  by the smooth red lines  (similar to the results of 
Ref. \cite{Ramaswamy}). The green dashed line refers to the time-reversed exponent
$^{(-)}\Lambda_3^{\mbox{GS}}$ and clearly emphasizes  the lack of any time-reversal 
symmetry  as expressed by the inequality (\ref{not_equal}). On the other hand,
for the covariant exponents precisely this symmetry is evident from the lower panel of Fig. \ref{figure_11}.  

To avoid confusion, we note that the `detailed balance symmetry' introduced in Ref. \cite{Ramaswamy}
is not connected with the symplectic local pairing symmetry considered here. The former only means
that for a global exponent to become zero, the positive and negative parts of the respective 
local exponent along the trajectory must cancel each other when integrated over time.   


\section{Concluding remarks}
\label{conclude}
  For the doubly-thermostated oscillator in a nonequilibrium stationary state, there is  a single vanishing 
global exponent, $\lambda_2$,  due to the time-translation invariance of the  equations of motion.
The corresponding covariant vector, ${\vec v}^2(t)$, needs to be parallel (or antiparallel) to the 
phase-space velocity $\dot{\vec \Gamma}(t) \equiv \{\dot{q}(t), \dot{p}(t), \dot{z}(t), \dot{x}(t)\}$. We have verified in our simulation that this is indeed the case.  The remaining vectors ${\vec v}^1$, ${\vec v}^3$ and $ {\vec v}^4$ are oriented with angles fluctuating between $0$ and $\pi$ with respect to  $\dot{\vec \Gamma}(t)$. 
The Gram-Schmidt vectors  behave very differently. Whereas the vector ${\vec g}^1$ is
identical to ${\vec v}^1$, the vector  ${\vec g}^2$ is not parallel
to   $\dot{\vec \Gamma}(t)$. Instead, the vectors ${\vec g}^3$ and  ${\vec g}^4$  are perpendicular
to   $\dot{\vec \Gamma}(t)$ as expected in view of the covariant  subspaces 
of  Eq. (\ref{subspaces}). These observations serve as convenient consistency checks for the
numerical procedure.

One of the remarkable features of the covariant  local Lyapunov exponents 
$\Lambda^{\mbox{cov}}({\vec \Gamma}(t))$ 
is their singular behavior transverse to the phase flow, whereas they are absolutely continuous in the direction 
of the flow. Fig. \ref{figure_9} provides an illuminating example.
The singularities are  consequences of bifurcations in the past history. Still, the local exponents 
are point functions in the phase space in the sense that one always gets the same value at the 
state point in question, as long as the trajectory has been followed from far enough back and has
experienced the same history. Due to the uniqueness of the solutions of  differential equations
there is only this path to the state point in question. The global exponents, however, are 
time averages of the local exponents along an (ergodic) trajectory.

    A final remark concerns the doubly-thermostated driven  oscillator again.  In a driven system
 (in our case a single particle in a non-homogenous thermal field)   
 heat and, hence, entropy is generated, which needs to be compensated by a negative entropy
 production in the thermostat to achieve a stationary state. The excess heat is transferred 
 from the system to the thermostat (by the positive friction  $z p > 0$), where it disappears.      
It follows from the thermostated motion equations in Sec. \ref{model_description} that the external 
entropy production (of the reservoir) is given by
$$
      \dot{S} / k  \equiv \frac{ \partial }{\partial {\vec \Gamma}} \cdot  \dot{\vec \Gamma}
          = z + x,
$$          
where $k$ is the Boltzmann constant.
In the non-equilibrium situation, a full time average $\langle z+x \rangle$  is necessarily positive. 
However, we have verified by simulation that finite time averages of this quantity numerically 
obey the steady-state  fluctuation theorem originally discovered by Evans, Cohen and Morriss \cite{ECM93}.
This theorem was given a firm theoretical basis by Gallavotti and Cohen \cite{GC95a,GC95b}, by
invoking the so-called `chaotic hypothesis' for Anosov-like systems.  Although our system is
not Anosov-like, it still obeys the theorem.

\section{Acknowledgements}  We gratefully acknowledge stimulating discussions with 
Francesco Ginelli, Josef Hofbauer, Gary Morriss, Antonio Politi, and G\"unter Radons.
Our work was supported by the FWF (Austrian Science Fund) grant P 18798-N20.

\end{document}